\PassOptionsToPackage{prologue,table,xcdraw}{xcolor}
\documentclass[letterpaper,twocolumn,10pt]{article}
\usepackage{arxiv}
\usepackage{tikz}
\usepackage{amsmath}

\usepackage{filecontents}
\usepackage{soul}
\usepackage{url}
\usepackage{booktabs}
\usepackage{multirow}
\usepackage{array}
\newcolumntype{L}{>{\raggedright\arraybackslash}X}
\usepackage{pdflscape}
\usepackage{makecell}
\usepackage{tikz}
\usepackage{xurl}
\usepackage{tabularx}
\usepackage{caption}
\usepackage{subcaption}
\usetikzlibrary{arrows.meta}
\usepackage{breakcites}
\usepackage{dirtytalk}
\usepackage{longtable}
\usepackage{makecell}
\usepackage{enumitem}

\NewDocumentCommand{\rot}{O{90} O{1em} m}{\makebox[#2][l]{\rotatebox{#1}{#3}}}

\AtBeginDocument{%
  \providecommand\BibTeX{{%
    \normalfont B\kern-0.5em{\scshape i\kern-0.25em b}\kern-0.8em\TeX}}}

\author{
  {\rm Naman Awasthi}\\
  University Of Maryland, College Park
  \and
  {\rm Saad Abrar}\\
  University Of Maryland, College Park
  \and
  {\rm Daniel Smolyak}\\
  University Of Maryland, College Park
  \and
  {\rm Vanessa Frias-Martinez}\\
  University Of Maryland, College Park
}

\newcommand{\plainauthor}{Author Name}
\begin{document}


\title{From "I have nothing to hide" to "It looks like stalking": Measuring Americans' Level of Comfort with Individual Mobility Features Extracted from Location Data}
\maketitle

\begin{abstract}
Location data collection has become widespread with smart phones becoming ubiquitous. Smart phone apps often collect precise location data from users by offering \textit{free} services, and then monetize it for advertising and marketing purposes. 
While major tech companies only sell aggregate behaviors for marketing purposes; data aggregators and data brokers offer access to individual location data. 
Some data brokers and aggregators have certain rules in place to preserve privacy; and the FTC has also started to vigorously regulate consumer privacy for location data. 
In this paper, we present an in-depth exploration of U.S. privacy perceptions with respect to specific location features derivable from data made available by location data brokers and aggregators. These results can provide policy implications that could assist organizations like the FTC in defining clear access rules. 
Using a factorial vignette survey, we collected responses from 1,405 participants to evaluate their level of comfort with sharing different types of location features, including individual trajectory data and visits to points of interest, available for purchase from data brokers worldwide. Our results show that trajectory-related features are associated with higher privacy concerns, that some data broker based obfuscation practices increase levels of comfort, and that race, ethnicity and education have an effect on data sharing privacy perceptions.We also model the privacy perceptions of people as a predictive task with F1 score \textbf{0.6}.
\end{abstract}



\section{Introduction}
Location (GPS) data collection has become widespread with smart phones becoming ubiquitous. Smart phone apps often collect precise location data from users by offering \textit{free} services like navigating using Maps, tracking fitness goals, finding relevant and customized search results, social networking and meetups, or location tracking for personal safety in times of crises, among others. 
Location data is used by app developers to understand app usage, improve functionalities and most importantly, to serve personalized ads and drive monetization. 

Monetization is driven by two distinct approaches. Tech companies like Google\footnote{\url{https://policies.google.com/technologies/location-data?hl=en-US\#how-is-location-information-used-for-ads}}, Apple or Meta collect precise location data from their users and use the data internally to improve app functionality while selling aggregate behaviors (\textit{e.g.,} geographical areas visited) for advertising purposes.
However, an individual's location data is never shared with or sold directly to third parties. On the other hand, data aggregators and data brokers collect, repackage, and sell individual location data to third-parties, including private organizations and academic researchers \cite{walkingActivity,hurricane_trd,homeworklocation,home_work_poi_pipeline}.
These data aggregators either develop their own apps, or create Software Development Kits (SDK) that are then shared with app developers to collect location data from specific apps. 
Well-known tech companies have faced lawsuits for location tracking without the user's consents \footnote{\url{https://www.tzlegal.com/news/37-5-million-settlement-facebook-tracking-users-location-without-consent/}},\footnote{\url{https://www.reuters.com/legal/google-pay-155-million-settlements-over-location-tracking-2023-09-15}}. 
The Federal Trade Commission (FTC) has also penalized two data brokers in 2024, for selling location data to advertisers without adequately informing consumers or obtaining their consent \cite{FTCcase}.   
However, while tech companies never share individual location data with third parties, data brokers do, and this has created several instances of inappropriate uses including the US Defense department's access to prayer app's location data to monitor Muslim communities  \cite{brennen_loophole}, local law enforcement agencies tracking racial justice protesters using location apps \cite{brennen_loophole}, and the use of location data purchased from data brokers to track gay priests \cite{grindr} and people who visit abortion clinics \cite{plannedparenthood}.


As a result of these news, some data aggregators and data brokers have created policies to prevent sensitive information about individuals being leaked. For example, data aggregator companies like Cuebiq don't allow the identification of visits to healthcare locations on their platforms, or only allow for the identification of home locations obsfuscated as census tracts \footnote{https://cuebiq.com/spoi-policy/}. 
However, although internal policy guidelines are important, they are solely based on data aggregators' decisions rather than on the perceptions of users whose location data is being monetized. We put forward that true consumer privacy rules need to come directly from consumer privacy perceptions.

In this paper, we conduct a thorough evaluation of user levels of comfort with respect to the use of individual location data - acquired from data aggregators and data brokers - to identify and characterize personal trips and visits to places. 
Studies in the past have used survey approaches to explore people's perceptions toward the collection and use of individual location data, with a focus on points of interest (POI) visited \cite{colombia_privacy_study,cross_platform_social_media_data,martin}. For example, past research has shown that users feel much more comfortable sharing their work or home location than their location when they attend a rally or visit a medical facility \cite{cross_platform_social_media_data}. 
However, points of interest are not the only feature available. In fact, travel history of millions of individual devices are available from data aggregators \footnote{\url{https://datarade.ai/data-categories/mobile-location-data}}. This location data is useful in deriving trajectory-based features such as origin-destination trips (\textit{e.g.,} trips from home to work) or the use of specific modes of transport (\textit{e.g.,} identifying driving vs. use of public transit vs walks).   
In this paper, we extend the state of the art in individual location data privacy perceptions by evaluating user perceptions with respect to trajectory-based features extracted from location data as well as points of interest, and we do so using U.S. representative surveys.



\textbf{Obfuscation Techniques.} 
As stated before, data aggregators have developed approaches to extract trajectory and visit features from location data and obfuscate details in privacy respectful ways. For example, instead of computing the exact home location, the location is computed as a census tract (SafeGraph);\cite{safegraph} or instead of sharing the exact route trajectory, data points are changed to obfuscate movement patterns towards sensitive locations and home/work location ( e.g., Data aggregators like spectus protects trajectories that end at home by replacing exact home location with the centroid of the census tract a person's home is in \cite{spectusDeviceRecurring}).
In this paper, we conduct an extensive review of current obfuscation approaches via published papers to identify the different types of techniques applied to individual location data acquired from data brokers; and we evaluate user privacy perceptions when using obfuscated location data vs. detailed (raw) location data. 
It is important to clarify that we exclusively focus on obfuscation techniques that directly change the location of a user as opposed to anonymization mechanisms that attempt to hide the actual location\cite{longroadcomp}.


\textbf{Factorial Vignette Approach.} Past literature like \cite{martin,ICADataCollection}, has shown that the level of comfort with location data sharing varies depending on who would be accessing the data and for what purpose. We follow Nissembaum's \cite{nissenbaum2004privacy} contextual integrity (CI) framework similar to \cite{martin} for creating survey questions that ask participants to evaluate their level of comfort with a specific feature to be used for a certain purpose by a certain actor. Using a factorial vignette survey approach we randomly create plausible combinations of actors, purposes and features to expose survey participants to different ways in which their location data could be used.
Understanding these nuanced privacy perceptions can support the creation of policies and rules concerning the types of features that can be extracted from location data, the types that should be avoided because they generate low levels of comfort, and the types of privacy respectful practices that are successful in increasing users' level of comfort. 
The FTC has started to vigorously regulate consumer privacy providing \textit{"bright-line rules"} for companies to clearly understand what can and cannot be done with location data \cite{FTCsWarn}.
This paper is an effort to illuminate this path, and provide guidelines for potential future FTC-proposed rules. To sum up, the main contributions and findings of this paper are: 

\begin{itemize}[leftmargin=*]

    \item   A thorough evaluation of user levels of comfort with respect to the use of trajectory data and Points of Interest taking into account actors and purposes. Our results show that people were uncomfortable particularly towards movement (trajectory-based) features such as frequent travel paths and were more comfortable with obfuscated home, work, and walk features. To the best of our knowledge, we are the first ones to reveal user perceptions with respect to the use of trajectory data. 
    
    \item An evaluation of users' level of comfort with respect to current data brokers' practices around the use of privacy respectful approaches to compute features from location data. Our results reveal a distinct preference for obfuscation approaches, especially when applied to detailed trajectory data. To the best of our knowledge, we are the first ones to evaluate user perceptions with respect to current data aggregator practices. 
    
    \item An analysis of how levels of comfort change with respect to educational background as well as racial and ethnic groups. Our analysis shows that there are significant differences in comfort between the racial groups White, Asian, Hispanic and Black. Additionally, participants with higher self-reported knowledge of computer science/programming were associated with higher comfort in sharing location features.    
     
\end{itemize}

\section{Literature Review}

\subsection{Location Data Tracking}
Location data can be collected using GPS sensors/Bluetooth beacons on mobile phone devices or smart-watches and using WiFi or IP addresses when surfing the web. Current app development toolkits offer the possibility of combining location data from multiple sensors to get better accuracy: for example, the Fused location Provider API \cite{googleFusedLocation} in Android development or the CLLocationManager \cite{appleCCLocation} in iOS development. 
In addition, there exist several third-party Software development Kits (SDKs) (e.g., X-Mode, SafeGraph) that help app developers, tech companies and data brokers and aggregators collect, visualize and track location data \cite{vox_privacy}.  

With data from data brokers and aggregators, 
many researchers have shown that location data characterizing the places visited by a person as well as their trajectories can be useful in a plethora of settings such as modeling the spreading of a pandemic through tracing apps \cite{contact_tracing_apps}, 
inferring socio-economic indicators \cite{hong2016topic,estabaan_urban_dynamics}, health outcomes \cite{garcia_you_are_what_you_eat}, urban development and disaster risk \cite{equalizer_social_infra,hurricane_trd}, 
targeted marketing and advertising \cite{location_advertising},
understanding migration patterns \cite{hong2019characterization},
improving public transit \cite{ma2014development}, 
or modeling mobility patterns for public safety \cite{wu2022enhancing}.

At the same time, there has been a lot of work focused on privacy preserving methods for location data. 
Primault et. al. \cite{longroadcomp} extensively outline the techniques for sharing location data in a privacy preserving manner either via anonymization or obfuscation approaches. While
obfuscation techniques directly change the location of a user to protect their privacy, anonymization mechanisms attempt to hide the actual location from an attacker. 
In this paper, we focus on obfuscation techniques, not on k-anonymization approaches, since obfuscation is the most common method used by data brokers, data aggregators and the researchers and analysts acquiring and analyzing location data from these companies.


\subsection{Privacy Perceptions for Location data}
Duckham et. al. \cite{ducham} define location privacy as "special type of information
 privacy which concerns the claim of individuals to determine for themselves when,
 how, and to what extent location information about them is communicated to others".
 Our research builds on prior work investigating privacy perceptions around sharing personal location data in various contexts \cite{martin,colombia_privacy_study,ICADataCollection}. Prior research has looked into the effects of different variables on the willingness of individuals to share their personal location data including the role of (1) Actors, defined as individuals accessing the location data \cite{martin,ICADataCollection,cross_platform_social_media_data,employee_surveillance} (2) Purposes, defined as how the location data is going to be used \cite{martin, mobile_data_thesis, cross_platform_social_media_data} (3) Time, defined as the duration of the access to location data \cite{martin,mobile_data_thesis} (4) Age  \cite{colombia_privacy_study,location_privacy_students,privacy_older_adults} (5) Privacy attitudes \cite{personality_location,personality_privacy_perception_2016}, and  (6) Socio-cultural aspects \cite{location_privacy_SEM,mobile_data_thesis,privacy_covid,surveillance_covid}.

These works have shown that actors like the FBI, commercial entities, employers and data brokers are negatively perceived while local government and social media companies generate more positive perceptions when collecting and using location data. The duration of the location data collection as well as the purposes of using location data have also been associated with varying levels of comfort when sharing that data. For example, while the identification of sexual orientation or political views are perceived negatively, purposes related to public health or public safety are perceived more positively.



There is also research exploring the use of location data for specific settings such as surveillance  \cite{video_survellience,vox_privacy}, location tracking for advertisements \cite{location_advertisement}, 
or location tracking for pandemic contexts. 
Surveillance applications have been associated with some level of comfort in human-centered contexts like disaster recovery.
On the other hand, research during COVID-19 found strong opposition towards contact tracing apps \cite{privacy_covid,surveillance_covid,privacy_covid_2}, with over 50\% of the participants in both studies being unlikely to install such apps due to concerns around government intrusion. 

Finally, prior work has also examined how monetary incentives influence participants' willingness to share their location data for specialized research studies or market research \cite{incentive_participate_1, incentive_participate_2}; and the relationship between personality traits and privacy perceptions \cite{personality_location, personality_privacy_perception_2016}. Participants' positive perception of the service collecting their data, higher incentives, agreeableness, conscientiousness, and openness to new experiences were also linked positively to comfort in sharing location data.  

Building on these findings, we extend the state of the art by (1) evaluating user perceptions with respect to trajectory-based features extracted from location data, beyond current analyses focused on visits to points of interest (POI), (2) evaluating user privacy perceptions when using obfuscated location data vs. detailed (raw) location data, and (3) analyzing the effect of race, ethnicity and education on privacy perceptions with respect to trajectory and POI visit location data. 

\subsection{Predicting Privacy Preferences}
Privacy fatigue is an important issue highlighted in research. This fatigue stems from factors like ubiquitous collection of personal data, scarcity of alternatives and additional permission requirements people need to go through \cite{tradeoff,privacy_exhaustion,privacy_risk_awareness}. 

Prediction models have been developed to explore people's perceptions around data collection and sharing like \cite{privacy_prediction,privacy_prediction2,privacy_prediction3,naeini,location_advertising,martin}. In these paper, the authors model people's perceptions of comfort in sharing location data as a multi-class classification or regression problem and model the perceptions around different actors accessing location data for different purposes. Eg: \cite{naeini} models people's perceptions in sharing IOT data as a binary classification task (predicting allow/deny); reporting accuracies of 77.6\% (0.70 recall) for a multi-class, 3-point Likert classification task, and 
81\% (0.78 recall) for the binary classification.
In another study, researchers created binary classifiers to determine privacy attitudes with regards to data sharing in several contexts including sharing biological data or personally identifying information (PII)\cite{contextualLabel}; achieving a mean recall of 0.46 for prediction of privacy concerns from contextual labels. 

In this paper, we explore binary and multi-class approaches, similarly to prior works.

\begin{table*}[ht!]
\centering
\footnotesize
\caption{Location Features and Abbreviations. We cite examples of recent studies where such features are used in the "feature cluster" column }
\begin{tabularx}{\textwidth}{|l|p{0.25\linewidth}|p{0.15\linewidth}|p{0.35\linewidth}|}
\cline{1-4} 
\multicolumn{1}{|l|}{Feature Cluster} & Feature & Abbreviation & Description \\ 
\cline{1-4} 
& Your inferred home location& Home location (Detailed)
& Home location as a point on map (Eg. fig. \ref{fig:homeD}) \\ \cline{2-4} 
 & Your inferred home location represented as   a census tract& Home location (Obfuscated)
& Home's county location (Eg., Fig. \ref{fig:homePP})\\ \cline{2-4} 
& Your inferred work location& Work location (Detailed)
& Work location as a point on map (Eg., Fig. \ref{fig:workD}) \\ \cline{2-4} 
\multirow{-4}{*}{ \shortstack[l]{Home + Work \cite{homeworklocation,home_work_poi_pipeline}}} & Your inferred work location represented as a census tract & Work location (Obfuscated)
& Work's county location (Eg., Fig. \ref{fig:workPP}) \\ 
\cline{1-4} 
& The places you visit & Places you visit (Detailed)
& Chart indicating types and frequency of places visited. Includes a map depicting the detailed locations of the places visited (Eg., Figs. \ref{fig:POVD}, \ref{fig:POVPP}) \\ \cline{2-4} 
& The types of places you visit & Places you visit (Obfuscated)
& Chart indicating types and frequency of places visited (Eg., Fig. \ref{fig:POVPP})\\ \cline{2-4} 
\multirow{-3}{*}{Places Visited\cite{PlacesofVisit,home_work_poi_pipeline,poi_2}}                     & The geographical area where you spend most   of your time & Area you spent most of your time (Obfuscated)
& Map displaying radius of gyration (Eg., Fig. \ref{fig:ROG})\\ \cline{1-4} 

& The modes of transportation you use, with what frequency and their corresponding routes & Modes of transportation (Detailed)
& Chart indicating frequency and types of mode of transport used. It also includes a map with lines indicating detailed routes for each frequent mode of transport (Eg., Fig. \ref{fig:modesD}, \ref{fig:modesPP}) \\ \cline{2-4} 
\multirow{-2}{*}{Transportation \cite{modesOfTransport_1,transport_NZ,home_work_poi_pipeline} }& The modes of transportation you use and   with what frequency & Modes of transportation (Obfuscated)
& Chart indicating frequency and types of mode of transport used (Eg., fig \ref{fig:modesPP})\\ \cline{1-4} 

& Your most frequent trips & Most frequent trips (Detailed)
& Map with frequently taken routes. Routes are detailed showing start and end locations as well as the in-between GPS points visited (Eg., Fig. \ref{fig:freqTD}) \\ \cline{2-4} 
& Your least frequent trips& Least frequent trips (Detailed)
& Map with infrequently taken routes. Routes show start and end locations as well as all the GPS points visited (Eg., Fig. \ref{fig:leastFreqT})\\ \cline{2-4} 
& Your most frequent types of trips & Most frequent type of trips (Detailed)
& Map with the different frequent trips taken and the inferred trip purpose by type of destination (Eg., Fig. \ref{fig:freqtypetrip})\\ \cline{2-4} 
& Your most frequent trips represented by their origin and destination census tracts and connected by a line& Most frequent trips between counties (Obfuscated)
& Map with frequently taken routes. It protects privacy by showing the start and end points as counties, and the frequently taken routes as straight lines between counties instead of detailed GPS. (Eg., Fig. \ref{fig:FreqOD}) \\ \cline{2-4} 
\multirow{-5}{*}{Movement \cite{modesOfTransport_1,traj_analysis,poi_2}} & Your most frequent trips represented by their origin and destination census tracts and connected by an approximate route & Most frequent trips between counties ('Google') (Obfuscated)
& Map with frequently taken routes. It protects privacy by showing the start points, end points as counties, and the frequently taken routes as suggested by  Google Maps, instead of the actual GPS route. (Eg., Fig. \ref{fig:gmaps})           \\ \cline{1-4} 

& Your walking activity and the corresponding routes & Frequent walking activity (Detailed)
& Chart indicating frequency and duration of walks. It also includes a map with detailed GPS trajectories indicating routes for each frequent walking path (Eg., Fig. \ref{fig:walkD} and \ref{fig:walkPP})\\ \cline{2-4} 
\multirow{-2}{*}{Walking Activity \cite{walkingActivity,mode_detection_all,transport_NZ,walk_nz}}                   & Your walking activity& Frequent walking activity (Ob)& Chart indicating frequency and duration of walks with no detailed trajectories  (Eg., Fig. \ref{fig:walkPP})                          \\ \cline{1-4} 

& The foreign countries you have visited and   the duration of the visit, including all the locations where you have stayed    & International visits (Detailed)
& Chart indicating frequency,duration and location of international trips. It includes a map with detailed GPS points indicating regions visited in the foreign county (Eg., Fig. \ref{fig:internationalD})\\ \cline{2-4} 
\multirow{-2}{*}{International Trips \cite{cuebiqInternational} }                & The foreign countries you have visited, and for how long& International visits (Ob)& Chart indicating frequency, duration and location of international trips. (Eg., Fig. \ref{fig:internationalP})                          \\ \cline{1-4} 

\end{tabularx}
\label{tab:listOfFeatures}
\end{table*}

\section{Survey Design}
Past work has shown that the level of comfort when sharing highly private information is influenced by \textit{who} accesses the data and for \textit{what} purpose, as stated by Nissembaum's contextual integrity framework \cite{nissenbaum2004privacy}. Hence, we used a factorial vignette survey approach where each vignette in the survey was created following the same format: \textit{Actor X wants to do Purpose Y and for that, they need to use Feature Z} while systematically changing actors, features and purposes across participants to test their relevance. For example, a potential survey vignette could be "A doctor wants to monitor your personal wellness. For that purpose they want to access your detailed walking activity. How comfortable would you feel with this use of your personal location data?". Levels of comfort were measured using a 5-point Likert scale, from \textit{Very Uncomfortable} to \textit{Very Comfortable}. 
Participants were also asked to fill out a free-form text box explaining their answer (see Figs. \ref{fig:locationdataqsdetailed} and \ref{fig:locationdataqspp} in the Appendix for two survey question examples).
Next, we describe location features, actors and purposes in detail. 

\textbf{Location Features}
We advance the state of the art in the evaluation of location data privacy perceptions along two main fronts. First, we consider features extracted from location data that go beyond current analyses focused on visits to points of interest (POI) such as visiting a hospital or a liquor store.
Specifically, we evaluate \textit{trajectory}-based features that are extracted from location data by considering sets of GPS points that define spatio-temporal datasets characterizing trajectories. 
For example, we can ask a participant about their level of comfort with a company using their detailed driving trips (showing the specific route followed) or their detailed walking activity.

Second, we evaluate trajectory-based and POI visit features using two approaches: detailed and obfuscated. 
While \textit{detailed} uses all GPS points available, \textit{obfuscating} approaches characterize location data using current state-of-the-art data aggregator practices of obfuscation. For example, an obfuscated driving trip would only specify the origin and destination census tract of the trip instead of the detailed trajectory with all locations (GPS points) visited in-between. Similarly, an obfuscated home location would only specify home location at the county level instead of the exact location.

Table \ref{tab:listOfFeatures} shows all the features we have used to craft the survey vignettes. 
These features can be categorized into six groups: "Home+Work", "Places Visited", "Transportation", "Movement", "Walking Activity", and "International Trips". 
For each feature in each category, we define both its Detailed (D) and its Obfuscated (Ob) version. 
Next to each feature in the Table, we provide links to papers that compute these features using 
location data acquired from data brokers or aggregators, showing that these are in fact state-of-the-art features when working with 
datasets from data aggregators. 
While POI features such as home, work and places visited have already been explored in the literature, comfort perceptions for obfuscating approaches for these have not been explored; nor the other trajectory features in either detailed or obfuscated forms. 
Each of these features is presented to the survey participants via an accompanying visualization. See the Description column in Table \ref{tab:listOfFeatures} for links to sample visualizations shown to survey participants. More details about survey design choices, including visualizations, are explained at the end of this section. 

\begin{table}[ht!]
\centering
\footnotesize
\caption{Actors and the abbreviations. Actors taken from previous studies like \cite{martin, mobile_data_thesis, ICADataCollection}}
\begin{tabular}{|p{0.55\linewidth}|l|}
        \hline
Actor & Abbreviation      \\
        \hline
{ An emergency service - like   emergency medical services or fire and rescue services } & Emergency services
\\
        \hline
    
A federal government agency - like the   FBI or CIA & Federal government agency
\\
        \hline
A local government agency & Local government agency
\\
        \hline
Your employer& Employer
\\
        \hline
A commercial entity & Commercial Entity\\
        \hline
Your family& Family\\
        \hline
Your doctor& Doctor\\
        \hline
A law enforcement agency - like a city police department or a county sheriff’s office & Law enforcement agency
\\
        \hline
Academic researchers& Academic researchers
\\
        \hline
\end{tabular}
\label{tab:listOfActors}
\end{table}

\textbf{Actors} 
Table \ref{tab:listOfActors} shows the list of the nine actors we use to create the vignette questions. 
These actors have been examined in previous studies related to location data privacy \cite{martin,ICADataCollection} and represent the majority of entity types that can access an individual's location information. 
Our work's novel contribution lies in looking at the effect of these actors on privacy perceptions in conjunction with novel trajectory-based features while controlling for demographic and educational data. 


\begin{table}[ht!]
\centering
\footnotesize
\caption{Purposes and Abbreviations. Purposes taken from previous studies like \cite{martin, mobile_data_thesis, ICADataCollection}}
\begin{tabular}{|p{0.6\linewidth}|p{0.3\linewidth}|}
\hline
Purpose   & Abbreviation               \\ \hline
{Understand criminal activity by looking   into the relationship between crime, people’s movements and locations visited} & Analysis of criminal activity
\\ \hline
Analyze terrorist attacks by looking into   people’s movements and locations visited                         & Analysis of terrorist attacks
\\ \hline
Monitor how people move (or don’t) to control the spread of a disease e.g., covid-19 & Control spread of diseases
\\ \hline
Monitor your personal wellness and physical activity & Personal wellness
\\ \hline
Show you targeted ads or personalized   announcements & Show Ads\\ \hline
Analyze traveling experiences and public transit services& Analysis of Public transit services
\\ \hline
Understand how people move in a city so as to inform the design of new walking and cycling infrastructure& Design new walking/cycling infrastructure
\\ \hline
Understand your commute to work so as to optimize work productivity & Optimize work productivity
\\ \hline
Monitor your mobility patterns e.g., the places you visit or the trips you make & Monitor mobility patterns
\\ \hline
Understand how people move so as to identify optimal locations for hospitals, libraries or parks& Identify locations for infrastructure
\\ \hline
\end{tabular}
\label{tab:listOfPurposes}
\end{table}
\textbf{Purposes}
Table \ref{tab:listOfPurposes} lists the ten different purposes we use when crafting the vignette questions. These purposes are informed by prior work looking into privacy perceptions related to location data  \cite{martin,ICADataCollection}, and cover public service purposes such as understanding where to build a new hospital, public health purposes such as monitoring population mobility during a pandemic, law enforcement such as using location data to identify criminal activity, as well as economic, commercial and general purpose such as using location data to show ads or optimize productivity.

There are combinations of actor, feature and purpose that do not make sense. For example, a doctor will not be interested in getting access to mobility patterns at city scale. Hence, we eliminate implausible combinations from the pool of possible questions. After this process, a total of 445 valid combinations of actor, purpose and feature are left and randomly shown to participants. 

We did a qualitative study with 5 respondents recruited from Craigslist to understand comprehension of vignette questions asked in the survey. They were asked to answer 10 randomly selected vignette questions and elaborate on their thought process. All the respondents were able to interact with the map visualizations, answer our questions about the contents of the map and did not raise any issues with the structure of the vignettes.


\subsection{Design Choices} 
Prior work in privacy risk perception studies has shown three important insights that we build on when creating the survey. 
First, \cite{privacy_risk_awareness} showed that abstract scenarios are often perceived as being less risky when compared to personal scenarios. Hence, before starting the survey questions, we ask participants to situate themselves in the vignettes as if this was their own personal data. In addition, the questions are phrased in a way that put the participant at the center of the vignette using the "you" pronoun
(see Appendix \ref{subsection:consentbrief} for more details on this design). 


Second, prior work has shown that visualizations can help better understand privacy risks, and that when privacy-related questions are asked with visualizations, privacy risk concerns with data sharing tend to decrease \cite{vizbetter}. Hence, each of the vignette questions in the survey is accompanied by an interactive visualization of the location feature, as well as by a brief explanation of that visualization to make sure that the user understand the feature extracted from the location data. Figure \ref{fig:locationdataqsdetailed} in the Appendix shows a visualization for a question where the feature is detailed trajectory data and transportation modes. The Appendix shows three other examples of feature visualizations including 
detailed features Frequent Trips (Fig. \ref{fig:freqTD}) and visits to points of interest (Fig. \ref{fig:POVD}) as well as an obfuscated feature showing frequent trips with synthetic trajectories generated with Google Maps (Fig. \ref{fig:gmaps}).


Third, prior work has shown that having prior technical knowledge or specific demographic or personality traits might affect privacy perceptions \cite{martin}. Hence, we also ask participants to fill out questions covering both privacy attitudes as well as technical knowledge and demographic data.
We broke these questions into two blocks, demographic and general tech knowledge questions were shown before the vignettes and the privacy attitudes questionnaire after. This design was based on user feedback from our qualitative survey evaluation, as it appeared to provide breaks before and after the more in-depth vignette questions. Following Martin et. al. \cite{martin}, the privacy attitudes questionnaire asked participants to rate on a 5-Likert scale (from "strongly disagree" to "strongly agree") their agreement with different privacy attitudes such as trust in business or in government (see Figs.\ref{fig:attitude_webpage} and \ref{fig:demographic_page} in the Appendix for a detailed list of all the questions asked, and section \ref{martinAppendix} for further details on the rationale behind the privacy questionnaire design).

\subsection{Platform} 
We advertised our online survey on Cint\footnote{\url{https://www.cint.com/}}. Interested participants clicked on a link on Cint that took them to an institution webpage where we hosted the survey. We did not use Cint surveys directly because of the elaborate nature of the factorial vignettes that were randomly sampled and of the custom map visualizations we created, consistent with the survey question. 
Each participant was paid \$7 for completing the survey which consisted of five vignette questions, seven demographic and computer knowledge questions, 10 privacy attitude questions. All participants were protected by IRB number 1768475-4 \footnote{\url{https://osf.io/x2vjk?view_only=e6dd3400991946afa3b265df5b1f3132}}. 

We collected a U.S. representative sample of 1,405 participants (7,025 vignette questions answered), which produced an average
of 16 answers per vignette, on par with prior work in privacy perceptions\cite{martin,ICADataCollection}.

\section{Analytical Approach}


As stated in the Introduction, the main objective of this paper is to analyze user levels of comfort with respect to the use of individual trajectory and POI visits taking into account the actors and purposes involved in the data analysis, as well as the presence (or not) of obfuscating approaches to preserve privacy. In addition, we are also interested in evaluating how these perceptions might change with respect the participants' educational or racial background. 

To carry out this analysis, we employ a mixed-effects ordinal logistic regression (clmm in R) due to the ordinal nature of the Likert responses (5-point Likert scale from (\textit{very uncomfortable}) to (\textit{very comfortable})). We follow best practices for the vignette data analysis using ordinal regressions \cite{vignette_instructions}. We use participant ID (each participant given a random ID) and vignette (vignette number) as two random effects. The dependent variable in the mixed effects ordinal regression is the level of comfort (five categories) and the independent variables are the variables whose effect we want to evaluate: actors, purposes, features (separating detailed features from features computed using obfuscation approaches), education and race and ethnicity. We also include the answers to the privacy attitudes, demographic and technical knowledge questionnaires as control variables in the regression, given that these have been associated with privacy comfort levels in prior work (\cite{martin}). 

A coefficient analyses of the independent variables will allow us to identify which variables are statistically significant and thus have an effect on the level of comfort that users have with different types of location data being shared. 
A detailed description of the ordinal regression proposed is provided in the Appendix, in section \ref{section:ordinal_regression}, where we also show that 
that assumptions for the ordinal regression are met (see Table \ref{tab:model_aic_compare} in Appendix).

In addition, we are also interested in evaluating the interaction effects between independent variables. 
For example, the ordinal regression might reveal a trajectory feature as having a significant negative effect on the level of comfort; but we are also interested in quantifying significant changes in levels of comfort for that feature across different types of actors or purposes. 
To run these analyses, we first transform the 5-point Likert scale categorical answers into a numerical scale from -2 to +2, with -2 being \textit{very uncomfortable}, 0 being a neutral privacy perception, and +2 being a \textit{very comfortable} level of comfort; and then 
run Kruskal-Wallis (KW) statistical tests between the pairs of variables whose interactions we want to evaluate (features and actors in our example); followed by post-hoc Dunn tests with the Benjamini-Hochberg correction to identify statistically significant differences in comfort levels across specific pairs (of features and actors in our example). 
Further details for the KW and post-hoc Dunn tests are provided in the Appendix, in section \ref{section:KWDunn}.
We do not run the interaction effects analyses within the ordinal regression because that would only allow us to explore relationships with respect to specific regression baselines, instead of across all feature values. 

In the next section, we provide some general statistics about the survey participants and the general distribution of levels of comfort across features, actors, and purposes. Next, in the following sections, we will discuss results for our three main research questions: 1) RQ1: effect of individual trajectory and POI visits on user levels of comfort, and their interactions with actors and purposes, 
2) RQ2: effect of obfuscating approaches on user levels of comfort, and 3) RQ3: effect of educational background and race and ethnicity on privacy perceptions. 

\section{Survey Response Analysis}
We collected answers from 1,405 participants. Each participant spent a mean of 18 min on the survey answering the 5 factorial vignette questions, as well as the demographic, computer knowledge and privacy attitudes questionnaires. 
From the 7,025 vignette questions answered, we removed answers whose required free-form text-box explanations did not clearly match the user comfort selected on the Likert scale, including both lack of quality explanations or low quality text. To achieve that, the answers were manually reviewed by two researchers who agreed upon its quality. 
In addition, we also removed answers with response times lower than 10 sec (given that the average adult reading speed is between 100-1200 words/min\cite{readingspeed} and the average vignette length is 43 words). 

We aim for representative sampling (proportions similar to U.S. Census from ACS \cite{census_education_attainment}. Survey population and ACS statistics are in Table \ref{tab:ethXeducensus}.

Table \ref{tab:ethXeducensus} shows the demographic characteristics across all survey participants for education levels as well as racial and ethnic groups, which are the focus of our analysis. (Please see Appendix for Age: Table \ref{tab:agedistribution} and Gender: Table \ref{tab:genderdistribution}). 
As can be seen, the sample is approximately representative of the U.S. population. 
The percentages for 
White and Black groups are close to ACS estimates, while the values for Asian and Hispanic are a bit lower. To account for this slight imbalance, we apply weights to our ordinal regression analysis, giving more importance to groups that are less represented (further details are provided in the Appendix \ref{section:ordinal_regression}).
%


\begin{table}[]
\centering
\footnotesize
\caption{U.S. census and survey population distribution across four major races/ethnicities and education levels.}
\begin{tabularx}{0.5\textwidth}{|l|XXX|X|}
\hline
\textbf{Education}   & \textbf{Under High School} & \textbf{High school to Bachelors} & \textbf{Bachelors and above} & \textbf{U.S. Census} \\
\textbf{Race/Eth} &&&& \\
\hline
\textbf{White}       & 6.2\%           & 37.4\%                  & 24.5\%                       & \textbf{68.2\%}      \\
\textbf{Black}       & 1.2\%           & 7.2\%                   & 2.9\%                        & \textbf{11.3\%}      \\
\textbf{Asian}       & 1.2\%           & 2.0\%                   & 3.2\%                        & \textbf{5.7\%}       \\
\textbf{Hispanic}    & 3.6\%            & 8.4\%                   & 2.9\%                        & \textbf{14.9\%}      \\
\hline
\textbf{Census} & \textbf{12.1\%}           & \textbf{55.0\%}                   & \textbf{33.5\%}              &                      \\
\hline
&                           &                                   &                              &\\
\textbf{Education}  
&           \textbf{Under High school}                &   \textbf{High school to Bachelors}                                &       \textbf{Bachelors and above}                       & \textbf{Survey Population}        \\
\textbf{Race/Eth} &&&& \\
\hline
\textbf{White}       & 3.60\%                    & 48.80\%                           & 21.40\%                      & \textbf{73.80\%}     \\
\textbf{Black}       & 1.00\%                    & 8.80\%                            & 3.30\%                       & \textbf{13.10\%}     \\
\textbf{Asian}       & 0.20\%                    & 2.80\%                            & 2.60\%                       & \textbf{5.60\%}      \\
\textbf{Hispanic}    & 0.40\%                    & 5.50\%                            & 1.70\%                       & \textbf{7.60\%}      \\
\hline
\textbf{Survey}        & \textbf{5.20\%}           & \textbf{65.90\%}                  & \textbf{29.00\%}             &  \\
\hline
\end{tabularx}
\label{tab:ethXeducensus}
\end{table}


\begin{figure}[ht]
    \centering
\includegraphics[width=0.5\textwidth]{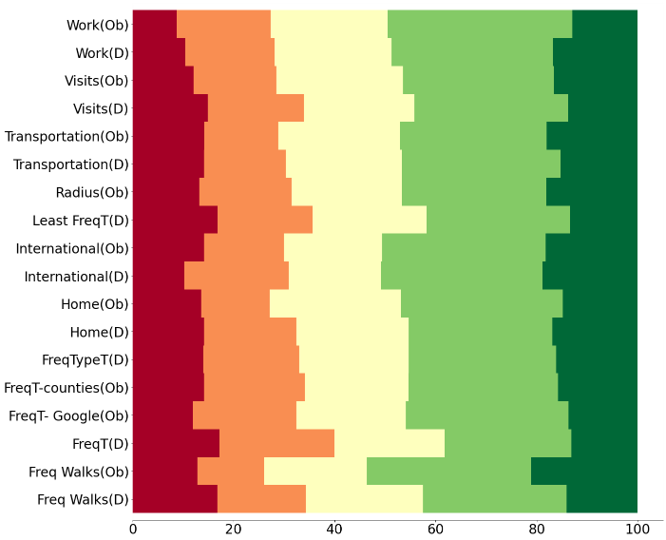}
    \caption{Percentage of responses per feature type and level of comfort (ordered left to right: Very uncomfortable, Uncomfortable, Neutral, Comfortable, Very comfortable). }
    \label{fig:features_responses}
\end{figure}

\textbf{Response Analyses.} Figure \ref{fig:features_responses} presents the distribution of participants' comfort levels for each location feature in our study. Trajectory features, including detailed most frequent trips, least frequent trips, most frequent walking trips, and places visited, were frequently labeled as \textit{Uncomfortable} or \textit{Very Uncomfortable}. 
In contrast, several obfuscated features showed considerably higher levels of comfort when compared to their detailed counterpart including frequent trips, frequent walking activity, or work location, among others. 
These numbers reveal that efforts to obfuscate individual trajectory and POI visit features are favorably seen by participants, increasing their comfort with these types of analytical approaches. 
Table \ref{tab:featcomfortmean} in the Appendix shows all mean comfort levels per location feature when levels are
transformed into a numeric variable. 


The privacy attitudes questionnaire at the end of the survey asked participants to rate their agreement with different statements related to trust in business or in government, or the role of authority, among others. 
Interestingly, trust in institutions and a general predisposition towards compliance appear to increase comfort levels with sharing location data (detailed or obfuscated), while skepticism towards authority is associated with reduced willingness to share location information. Detailed results can be explored in the Appendix (see Table  \ref{tab:attitudes_responses}). These results confirm the importance of adding participant responses to privacy attitude questionnaires in the regression model as control variables, since they appear to play a role in the perception of privacy comfort.

The distribution of comfort levels for each purpose and for each actor considered in the study confirmed prior work \cite{martin,CISmartHome,contact_tracing_apps,ICADataCollection}
 (see Figure \ref{fig:actor_responses} and Table \ref{tab:actorcomfortmean} in the Appendix for detailed results). 
Participants more frequently responded with "Very Uncomfortable" or "Uncomfortable" for purposes related to optimizing work productivity and monitoring mobility patterns; 
while purposes with perceived societal benefits, such as identifying locations for infrastructure development, designing walking/cycling infrastructure, and analyzing public transit services, were more frequently rated as \textit{comfortable} or \textit{very comfortable}.
On the other hand, and also confirming prior work, participants expressed greater discomfort sharing data with Employers and Federal Government Agencies; while actors such as Academic Researchers, Family, Emergency Services, and Doctors were rated higher on the comfort scale (see Figure \ref{fig:purpose_responses} and Table \ref{tab:purposecomfortmean} in the Appendix for further details).
\begin{table*}[!htbp]
\centering
\footnotesize
\caption{Regression coefficients for Actors, Purposes, Features, Ethnicity and Education. Odds ratio column is the likelihood of being more comfortable compared to the baseline class. Coefficients are only show for actor, purpose, feature, ethnicity and education. Other control variables are shown in the Appendix Table \ref{tab:attitude_privacy_coeffs}}
\begin{tabularx}{\textwidth}{llllll}
   & Coefficient       & Odds Ratio & Std Err   & Statistic & p-value   \\
\textbf{Actor (Baseline: Commercial Entity)}& & & & & 
\\
Federal government agency&\textbf{ -0.249**}& 0.78& 0.11& -2.266& 0.023
\\
Law enforcement agency& 0.14& 1.15& 0.11& 1.272& 0.203
\\
Local government agency& 0.105& 1.111& 0.097& 1.081& 0.28
\\
Academic researchers& \textbf{0.539**}& 1.714& 0.087& 6.199& 0
\\
Emergency services&\textbf{ 1.166**}& 3.209& 0.185& 6.297& 0
\\
Doctor& \textbf{1.078**}& 2.939& 0.169& 6.393& 0
\\
Employer&\textbf{ -0.626**}& 0.535& 0.149& -4.205& 0
\\
Family&                  \textbf{ 1.409**}&            4.092&           0.166&           8.493&           0
\\
\textbf{Purpose (Baseline: Show ads)}& & & & & 
\\
Analysis of terrorist attacks& 0.09& 1.094& 0.183& 0.488& 0.626
\\
Analysis of Public transit services& -0.047& 0.954& 0.185& -0.255& 0.799
\\
Control spread of diseases& 0.138& 1.148& 0.188& 0.731& 0.465
\\
Monitor mobility patterns& \textbf{-0.654**}& 0.52& 0.178& -3.666& 0
\\
Personal wellness& -0.338& 0.713& 0.212& -1.594& 0.111
\\
Analysis of criminal activity& -0.075& 0.928& 0.179& -0.416& 0.677
\\
Design new walking/cycling infrastructure& \textbf{0.583**}& 1.791& 0.195& 2.986& 0.003
\\
Identify locations for infrastructure& 0.296& 1.344& 0.19& 1.56& 0.119
\\
Optimize work productivity& 0.152& 1.164& 0.224& 0.679& 0.497
\\
\textbf{Feature (Baseline: Places you Visit (Detailed))}& & & & & 
\\
International visits (Detailed)& -0.061& 0.941& 0.209& -0.29& 0.772
\\
International visits (Obfuscated)& \textbf{0.369*}& 1.446& 0.21& 1.753& 0.08
\\
Area you spent most of your time (Obfuscated)& 0.012& 1.012& 0.148& 0.081& 0.936
\\
Modes of transportation (Obfuscated)& 0.192& 1.212& 0.144& 1.327& 0.185
\\
Modes of transportation (Detailed)& 0.01& 1.01& 0.149& 0.066& 0.948
\\
Places you visit (Obfuscated)& 0.166& 1.181& 0.139& 1.191& 0.234
\\
Home location (Obfuscated)& \textbf{0.441**}& 1.554& 0.213& 2.068& 0.039
\\
Home location (Detailed)& -0.101& 0.904& 0.21& -0.48& 0.631
\\
Work location (Obfuscated)& \textbf{0.376**}& 1.456& 0.19& 1.973& 0.049
\\
Work location (Detailed)& -0.2& 0.819& 0.191& -1.043& 0.297
\\
Least frequent trips (Detailed)& -0.198& 0.82& 0.153& -1.292& 0.196
\\
Most frequent trips between counties (Obfuscated)& 0.168& 1.183& 0.148& 1.134& 0.257
\\
Most frequent trips between counties ('Google') (Obfuscated)& 0.23& 1.259& 0.147& 1.562& 0.118
\\
Most frequent trips (Detailed)& \textbf{-0.297**}& 0.743& 0.146& -2.039& 0.041
\\
Most frequent type of trips (Detailed)& -0.063& 0.939& 0.147& -0.431& 0.667
\\
Frequent walking activity (Detailed)& -0.191& 0.826& 0.143& -1.335& 0.182
\\
Frequent walking activity (Obfuscated)&  \textbf{ 0.458**}&            1.581&           0.145&           3.166&           0.002
\\
\textbf{Ethnicity/Race (Baseline: White)}& & & & & 
\\
Asian& -0.017& 0.983& 0.233& -0.072& 0.942
\\
Black& 0.221& 1.247& 0.162& 1.361& 0.174
\\
Hispanic&  \textbf{0.504**}&            1.655&           0.196&           2.574&           0.01
\\
\textbf{Education (Baseline: Highschool to Bachelors)}& & & & & 
\\
Bachelors and above& -0.206& 0.814& 0.126& -1.631& 0.103
\\
Under Highschool & -0.017 & 0.983 & 0.227 & -0.075 & 0.94\\
\multicolumn{6}{r}{\tiny Significance ** $p-value < 0.05$, * $p-value < 0.1$} 
\end{tabularx}
\label{tab:apf_reg}
\end{table*}

\section{RQ1: Effect of Location Features, Actors and Purposes on Privacy Perceptions}
Table \ref{tab:apf_reg} shows the ordinal regression coefficients and odds ratios across actors, purposes and location features. 
Statistically significant coefficients reveal features that have a significant effect on participants' levels of comfort.
The odds ratio indicates the odds of that effect being more or less likely with respect to the baseline, holding all other variables constant.   

First, we analyze the effect of location features (individual trajectory and POI visit data) on privacy perceptions, and evaluate interaction effects between location features and actors or purposes. 
Next, we describe our main findings for actors and purposes, confirming significant insights found in prior work. 

\subsection{Location Features}



Table \ref{tab:apf_reg} shows that participants expressed statistically significant lower levels of comfort when sharing detailed trajectory data compared to detailed visits to POIs (regression baseline). For example, the odds of being comfortable sharing Most Frequent Trips (Detailed) was 0.74 times that of sharing detailed visits to POIs. In other words, participants were significantly less comfortable with the use of detailed trajectory data than with the use of visits to specific types of POIs (restaurants, libraries, schools, etc).
As one participant stated:
"There is no good reason for them to monitor my trips on the regular at all."
(Participant-8a6a). 

Participants, on the other hand, demonstrated higher comfort levels 
sharing obfuscated trajectory data compared to detailed visits to POIs. For example, walking (Frequent Walking Activity) was 1.5 times more likely to be rated as comfortable and international trips (International Visits) were 1.44 times more likely to be rated as comfortable than detailed visits to specific points of interest. 
Finally, participants were significantly more comfortable sharing their home and work locations (obfuscated to the census tract) than sharing detailed visits to specific POIs, although no significant difference was observed when home and work locations were not obfuscated (we discuss this further in RQ2).
Interestingly, these results reflect that if handled appropriately, users are willing to share some of their trajectory data. As  
Participant-D303 shared, "This is general information and I would be interested in this as well". 

 \begin{table}[ht]
\centering
\footnotesize
\caption{Kruskal-Wallis analysis with post-hoc Dunn test results for a subset of pairs (full analysis in Table \ref{tab:kw_featureXactor_featureXpurpose_BIG} in Appendix). M"i" represents median level of comfort value for distribution "i". Significance at p-value <0.05)}
\begin{tabularx}{\linewidth}{|p{0.15\linewidth}|p{0.18\linewidth}|p{0.18\linewidth}|p{0.05\linewidth}|p{0.05\linewidth}|p{0.08\linewidth}|}
\hline
\textbf{Feature} & \textbf{Actor1} & \textbf{Actor2} & \textbf{M1} & \textbf{M2} & \text{\footnotesize{M2-M1}}\\
\hline

Freq Walks(Ob) & Employer & Researchers & 0 & 1 & 1 \\
\cline{3-6}
&& Doc & 0 & 1 & 1 \\
\cline{3-6}
 &  & Family & 0 & 1 & 1 \\
 \hline
Freq Walks(Ob) & Fed & Researchers & 0 & 1 & 1 \\
 \cline{3-6}
 &  & Doc & 0 & 1 & 1 \\
 \cline{3-6}
 & & Family & 0 & 1 & 1 \\
 \hline
Freq Walks(Ob) & Local gov & Doc & 0 & 1 & 1 \\
\hline
\textbf{Feature} & \textbf{Purpose1} & \textbf{Purpose2} & \textbf{M1} & \textbf{M2} & \text{\footnotesize{M2-M1}} \\
\hline
Visits(Ob) & Monitor mobility & Public transit & 0 & 1 & 1 \\
\cline{3-6}
&  & Infra (Walk/Cycling) & 0 & 1 & 1 \\
\hline

\end{tabularx}

\label{tab:kw_featureXactor_featureXpurpose}
\end{table}

\textbf{Interactions with Actors and Purposes}

Table \ref{tab:kw_featureXactor_featureXpurpose} shows a subset of results for the Kruskal-Wallis (KW) and Dunn tests between the levels of comfort for individual trajectory data and POI visit features and the different types of actors and purposes considered in our study (please see Table \ref{tab:kw_featureXactor_featureXpurpose_BIG} in the Appendix for full details). 
The results show two interesting findings. First, the  
median level of comfort sharing obfuscated trajectory data decreases depending on the actor accessing the datwhile for 
  researchers, family, or doctors the median level of comfort is "comfortable" (median=1, p-value<0.05) for Federal Government agency or Employer actors the level of comfort is neutral (median=0, p-value<0.05)).
Second, the levels of comfort with respect to visits to POIs (Places you visit (Ob)) has significantly higher median values for purposes with social impact such as "Identifying locations for infrastructure or cycling" (median=1, p-value<0.05) than for more generic purposes, such as "monitoring mobility patterns" (median=0, p-value<0.05).



\subsection{Actors and Purposes}

Our ordinal regression results confirm prior findings for the effect of actors and purposes on privacy perceptions \cite{martin, ICADataCollection}.  
Table \ref{tab:apf_reg} shows that 
Federal Government Agencies and Employers were significantly negatively related to comfort when accessing location data. In fact, participants had 0.78 and 0.53 times the odds (p<0.01) of rating their level of comfort lower for these actors than for commercial entities (regression baseline). 
On the other hand, Academic Researchers, Emergency Services, Family members, and Doctors were positively related to levels of comfort, with participants having
1.7, 3.2, 4 and 2.9 times the odds (p<0.01) of rating their level of comfort higher for these actors than for commercial entities (baseline).  

Moving on to purposes, we observe that purposes with social impact (e.g., Designing public infrastructure) were perceived more favorably than marketing purposes (regression baseline), with the odds ratio being 1.79 times higher. 
On the other hand, more general purposes (e.g., Monitor mobility patterns) were associated with significantly lower comfort levels compared to the baseline, i.e. people were approximately half as likely to be comfortable with sharing data (p-value<0.05).

\textbf{Interaction Analysis.} 
A Table with the Kruskal-Wallis and Dunn test results for the interaction between actors and location features can be found in Table \ref{tab:kw_actor_feature_BIG} in the Appendix. 
We would like to highlight a couple of significant findings. 
First, while the effect of the actor being a Doctor is positive on the level of comfort (as we showed in Table \ref{tab:apf_reg}), the perception is more positive when sharing walking activity (obfuscated) (median level of comfort=1) than other features such as most frequent type of trips (obfuscated) (median=0, neutral) or most frequent trips (detailed) (median=-0.5, leaning uncomfortable).
On the other hand, when the actor is an Employer, giving access to the most frequent places of visit (obfuscated) was perceived as more uncomfortable (median=-1, uncomfortable) than giving access to walking activity (obfuscated) (median=0, neutral). 
Finally, no significant differences were identified between purposes and location features, pointing to similar levels of comfort independently of the trajectory or POI visit feature.

\section{RQ2: Effect of Obfuscating Approaches on Privacy Perceptions}

Looking into the Features column in Table \ref{tab:apf_reg} we can observe that some obfuscated trajectory and POI visit features are associated with statistically significant higher levels of comfort than detailed features (Places you visit (detailed) as baseline).
Specifically, obfuscated frequent walking activity as well as obfuscated international trips, obfuscated home and work locations had
1.58, 1.4, 1.55 and 1.45 times the odds (p<0.05) of being rated higher in level of comfort than detailed places visited, 
pointing to individuals being more prone to sharing certain types of trajectory and visits data if protected by obfuscation approaches. 
Beyond statistical significance, it is important to highlight that when looking into mean levels of comfort for each detailed and obfuscated location feature, all means were higher for obfuscation features than for their detailed counterpart e.g., Home (Detailed - Obfuscated) = 0.055, Modes of Transport (Detailed - Obfuscated) = 0.045.   
In other words, although not significant in the regression model, obfuscated location features are associated with more positive levels of comfort towards data sharing. Table \ref{tab:featcomfortmean} in the Appendix shows all mean levels of comfort across location features. 

\textbf{Interaction Analysis.} 
We perform KW and post-hoc Dunn tests between the levels of comfort associated with "Detailed" location features and "Obfuscated" location features across types of actors and purposes. Detailed results are shown in 
the Appendix Table \ref{tab:kw_ap_obfuscate_BIG}. Here, we discuss two relevant findings.
First, we observe that when the actor accessing the location feature is a family member (actor=Family), the user level of comfort increases from neutral to comfortable if the location feature is obfuscated instead of detailed. 
Similarly, we found that obfuscated features improved the privacy level of comfort when compared to detailed features for certain purposes such as "control the spread of diseases". This finding aligns with studies that show privacy preserving location sharing apps during COVID were perceived more positively compared to detailed location sharing app\cite{privacy_covid,privacy_covid_2,surveillance_covid}.

\section{RQ3: Effect of Race/Ethnicity and Education on Privacy Perceptions}

\subsection{Race and Ethnicity}
The results in Table \ref{tab:apf_reg} show that Hispanic participants are associated with statistically significant higher levels of comfort that White participants (regression baseline), having 1.6 times the odds (OR=1.6, p<0.01) of rating their level of comfort with location data higher than White participants.  
Some of the free-form text boxes align with these observations. For example, participants who self-identified as Hispanic said: "I have nothing to hide so it doesn't really make a difference to me." (Participant-050e); and 
"(In) this particular scenario I feel very comfortable with my data being accessed \& used for this purpose. Hospital locations are important and their location within a community can severely impact the quality of life for that community too." (Participant-2ae1).
On the other hand, no significant difference in level of comfort was observed between any other racial or ethnic groups. 

\textbf{Interaction analysis.} The KW and Dunn statistical tests between the levels of comfort across race/ethnicity and location data types, actor types and purposes revealed interesting nuanced privacy perceptions. Detailed results can be found in Tables \ref{tab:kw_ethnicity_act,tab:kw_ethnicity_p_f_priv} in the Appendix. Here, we discuss the most relevant findings. For Hispanic individuals, the levels of comfort were the lowest when their location data was used for marketing purposes i.e. showing ads (median=-1 "uncomfortable, p-value<0.05). 
In contrast, Black participants perceived marketing purposes as positive (median=1 "comfortable", p-value< 0.05) when compared to other purposes like analysis of terrorist attacks (median=0), public transit (median=0), mobility monitoring (median=0).

Nevertheless, no significant differences were observed between Hispanic ethnicity and actor or feature types; neither between Black and actors or feature types.  
This might indicate that, for Hispanic and Black participants, data comfort revolves around purpose of data access, and not around who access the data (actor) or the specific type of data accessed (feature). 


On the other hand, for Asian and White respondents there was a significant change in level of comfort between detailed and obfuscated features with, for example, "Most frequent trips (Detailed)" having a lower median comfort (median=0, p-value<0.05) than "Frequent walking activity (Obfuscated)" (median=1) for White individuals. 
In addition, the tests also showed Asian and White respondents being significantly less comfortable with the Federal Government, Law enforcement, Local government (median=0) and Employers (median=-1) than with Researchers, Family, Doctors or Emergency services (median=1). 
White participants were also more comfortable in sharing data for infrastructure purposes (designing cycling infrastructure and designing built infrastructure) (median=1, p-value<0.05) than for all the other purposes (median=0).
Thus, for Asian and White participants we find evidence that the combinations of actors, purposes and features are responsible for the overall privacy perception.

\subsection{Education}
We do not observe any statistically significant difference in comfort levels for different education groups when compared to participants with education high school to bachelors (baseline). 

\textbf{Interaction analysis.} Kruskal-Wallis and post-hoc Dunn tests to assess interactions between levels of comfort across education levels and actors, purposes and location features revealed some interesting insights (see Table \ref{tab:kw_bachelors_apf} in the Appendix for full details). Participants with a "Bachelors and above" felt more comfortable in sharing obfuscated features (median=1,pval<0.05) compared to detailed (median=0, p-value<0.05). 

Interestingly, no statistical difference between detailed and obfuscated location data was observed for the other two educational groups "High School to Bachelors" and "Under High School". 
Participants with a "Bachelors and above" also showed significantly lower levels of comfort with purposes related to monitoring mobility (median=0, p-value<0.05) and with sharing location data with employers (median=-1, p-value<0.05). These findings indicate that decisions of this educational group depend on the interaction of the full spectrum of actor-purpose-feature (along with privacy attitudes). 

Participants with education "High School to Bachelors" were significantly less comfortable in sharing their location data with the Federal Government, law enforcement agencies (median=0, p-value<0.05) or employers (median=-1,p-value<0.05) than with Researchers and Emergency services (median= 1,pval<0.05). These results reveal that the privacy perceptions for this educational group are mostly shaped around actors and purposes, and not types of features or obfuscation techniques. Finally, we did not observe any significant difference between pairs of actors, purposes or features for participants with "Under High School" education, signaing that for these participants their decisions might be mostly shaped by their individual privacy attitudes.

\section{Predictive Modeling}
This section explores the creation of models that predict levels of comfort in sharing location data a given a combination of actor, purpose and location feature. 
Such models can help companies evaluate levels of comfort with different types of location features before they are actually implemented and deployed in their location intelligence platforms.  
For example, a company could be interested in evaluating the level of comfort of a Hispanic person in sharing their movement data with law enforcement for public safety.

\subsection{Model}
We frame the prediction task as a classification task with the classes being the levels of comfort. We explore two types of models. In \textit{Model 1}, our objective is to predict the level of comfort of a given demographic group with a given vignette (framed as actor, location feature, factor). This model would allow a company to evaluate the level of comfort with the use of a location feature associated with a specific population group.  
We also evaluate a second model \textit{Model 2}, whereby privacy attitudes from the population of interest are also known, either because the company knows them, or because they can run a privacy attitudes survey among their population of interest. 
 
We consider three different approaches to defining the classes in the classification task: 1) 5-point Likert scale (as used in the original survey), 2) 3-point Likert scale ((very) uncomfortable, neutral and (very)comfortable), and 3) Binary classification ((very) uncomfortable or neutral vs. (very) comfortable). We perform an 80\% train 20\%test split for training and model evaluation, and compare  the performance of these machine learning algorithms (suitable for classification tasks): Random Forest (RF), Gradient Boosting Machines (XGB) and K-Nearest Neighbours (KNN). To assess model performance, we utilize the following metrics:
Accuracy, Precision, Recall, F1-score and Cohen's Kappa.
 

\subsection{Performance Analysis}

\textbf{Model 1.} Table \ref{tab:pred_context} presents the performance metrics for predicting comfort levels given a tuple of actor, purpose and feature as well as general demographic information of the population of interest to the company. 
We can observe moderate accuracy and F1 scores across models and number of classes with ranges (0.299-0.695) and (0.279-0.325), respectively. As expected, higher number of classes are associated with lower accuracy values. Nevertheless, the F1 scores are the lowest for the binary classification pointing to challenges in balancing precision and recall, possibly due to class imbalance.
The moderate performance across all models suggests that vignette and demographic information alone may not be sufficient for highly accurate predictions of comfort levels in location data sharing. This underscores the complexity of privacy perceptions and the potential influence of other factors not captured in demographic data such as privacy attitudes.

\begin{table}[]
\centering
\footnotesize
\caption{Detailed classification report for the three labeling approaches. Dependent variables: Actor, purpose, feature, demographic attributes Best metric per classification approach are in bold. XGB: XGBoost, KNN: K-nearest Neighbours, RF: RandomForest, CK: Cohen Kappa score}
\begin{tabularx}{0.5\textwidth}{@{}lc*{5}{>{\centering\arraybackslash}X}@{}}
\toprule
\textbf{Model} & \textbf{Class} & \textbf{Acc.} & \textbf{Prec.} & \textbf{Rec.} & \textbf{F1} & \textbf{CK} \\
\midrule
XGB            & 2              & \textbf{0.695} & 0.155         & \textbf{0.455} & 0.232      & \textbf{0.095} \\
KNN            & 2              & 0.647          & 0.219         & 0.349          & 0.269      & 0.052 \\
RF             & 2              & 0.658          & \textbf{0.223} & 0.372         & \textbf{0.279} & 0.073 \\
\midrule
XGB            & 3              & \textbf{0.481} & 0.587         & 0.481          & 0.521      & \textbf{0.118} \\
KNN            & 3              & 0.478          & \textbf{0.605} & 0.478         & 0.527      & 0.106 \\
RF             & 3              & 0.459          & 0.530         & \textbf{0.459} & \textbf{0.486} & 0.096 \\
\midrule
XGB            & 5              & \textbf{0.299} & \textbf{0.385} & \textbf{0.299} & \textbf{0.325} & \textbf{0.060} \\
KNN            & 5              & 0.287          & 0.380         & 0.287          & 0.318      & 0.042 \\
RF             & 5              & 0.293          & 0.351         & 0.293          & 0.313      & \textbf{0.060} \\
\bottomrule
\end{tabularx}
\label{tab:pred_context}
\end{table}
\textbf{Model 2.} Here, we examine the predictive performance of Model 1 enhanced with privacy attitude data from the population of interest. 
As Table \ref{tab:pred_attitude_context} shows, adding privacy attitudes as independent variables in the prediction model consistently improves prediction performance across all metrics: accuracy, precision, recall, and Cohen-Kappa; with improvements over 50\% for 3-point and 5-point Likert scales when compared to Model 1. Accuracy and F1 score values for Model 2 were in the ranges (0.517,0.783) and (0.517,0.600), respectively. The best performing model was the Random Forest which showed superior performance across all classification settings (binary and multi-class).

\begin{table}[]
\centering
\footnotesize
\caption{Detailed classification report for the three labeling approaches. Dependent variable: Actor, purpose, feature, demographic attributes, privacy attitudes and trust. Best metric per classification approach are in bold. XGB: XGBoost, KNN: K-nearest Neighbours, RF: RandomForest, CK: Cohen Kappa score}
\begin{tabularx}{0.5\textwidth}{@{}lc*{6}{>{\centering\arraybackslash}X}@{}}
\toprule
\textbf{Model} & \textbf{Classes} & \textbf{Acc.} & \textbf{Prec.} & \textbf{Rec.} & \textbf{F1} & \textbf{CK} \\
\midrule
XGB            & 2                & 0.772         & 0.522          & 0.642         & 0.576       & 0.422       \\
KNN            & 2                & 0.780         & 0.522          & \textbf{0.665}& 0.585       & 0.439       \\
RF             & 2                & \textbf{0.783}& \textbf{0.549} & 0.661         & \textbf{0.600} & \textbf{0.452} \\
\midrule
XGB            & 3                & 0.621         & \textbf{0.649} & 0.621         & 0.632       & 0.386       \\
KNN            & 3                & 0.593         & 0.588          & 0.593         & 0.589       & 0.364       \\
RF             & 3                & \textbf{0.642}& 0.647          & \textbf{0.642}& \textbf{0.644} & \textbf{0.433} \\
\midrule
XGB            & 5                & 0.476         & 0.486          & 0.476         & 0.478       & 0.318       \\
KNN            & 5                & 0.464         & 0.462          & 0.464         & 0.462       & 0.314       \\
RF             & 5                & \textbf{0.517}& \textbf{0.520} & \textbf{0.517}& \textbf{0.517} & \textbf{0.375} \\
\bottomrule
\end{tabularx}
\label{tab:pred_attitude_context}
\end{table}

\section{Implications for Policy and Practice}
\textbf{Location Features.} Our regression analyses demonstrate that participants have higher levels of comfort in sharing detailed visits to POIs than detailed trajectory data. However, obfuscating some of the trajectory variables resulted in higher levels of comfort even above sharing detailed visits to POIs.
Our analyses also showed that these levels of comfort are also increased when the actors accessing the data are researchers or family, or when the purpose is focused on social good. 
These findings point to the implementation of more granular control mechanisms that allow users to specify which types of location features can be derived from the data collected from them. 
Since data brokers and data aggregators collect data from apps via SDKs, they could also allow users to select, during the app installation and when they are notified about the location data collection, which one of the six feature clusters shown in Table \ref{tab:listOfFeatures} they would be comfortable sharing their location data for. Users should also have an opt-out mechanism for which type of actor and purpose their data can be used for. While arguing against the purchase of location data from data brokers by government agencies, legal scholars and practitioners like Rahbar \cite{carpenterEvade} and Conner \footnote{\url{https://www.lawfaremedia.org/article/data-broker-sales-and-the-fourth-amendment}}, have raised serious concerns about such purchases violating the fourth amendment.


\textbf{Obfuscation Approaches.} Higher levels of comfort were clearly observed for some obfuscated location features (both trajectory and visits to POIs) when compared to detailed visits to POIs (baseline) in the regression model \ref{tab:apf_reg}. These levels of comfort were increased when the actor accessing the data was a family member. 
Although not all obfuscated approaches were significantly associated with higher levels of comfort in the regression, the mean values were lower. 
We posit that maybe there is no need for more granular control over obfuscated vs. detailed features. By default, the features could be obfuscated, and during the app installation users could be offered to opt-in to share more detailed features, together with an explanation of why this would be valuable for analytical purposes (actor, purpose).  
 In future research, we will also explore why obfuscation techniques appear to not always alleviate privacy concerns.




\textbf{Race, Ethnicity and Education Effects.} Our regression analysis has revealed significant differences in comfort levels for the Hispanic participants when compared to White participants, with Hispanic participants being more comfortable sharing trajectory and visit data. Our interaction analysis has also shown a complex interplay between 
cultural factors, educational experiences, and privacy perceptions while controlling for privacy attitudes and computer knowledge. Hispanic and Black participants as well as "High School to Bachelors" appear to shape their perceptions mostly on the purpose of the data access, or on who has access to the data (actor). On the other hand, White participants and participants with educational level "Bachelors and above" appear to shape their privacy perceptions around actors, purposes and by the types of location data being shared. 
These findings could be used to pre-populate some of the initial app selections for types of location data, actors or purposes based, for example, on the educational background of the person installing the app, thus reducing the burden of having to select among many different options from scratch. 



\section{Conclusion}
This study contributes to our understanding of the nuanced nature of privacy perceptions in location data sharing. By highlighting the importance of the type of location feature (trajectory or visits to POIs),
the presence of an obfuscation approach, contextual factors such as actors and purpose, attitudes and demographic variations, our findings provide a foundation for more user-centric, feature focused approach to location privacy. 
Our results have shown that trajectory-related features are associated with higher privacy concerns; that current obfuscation approaches by location data brokers and aggregators are sometimes successful; and that race, ethnicity and education have an effect on privacy perceptions with Hispanic and High School to Bachelor populations being associated with higher levels of comfort. 
As technology continues to evolve, ongoing research in this area will be crucial for developing privacy practices and policies that effectively balance the benefits of location-based services with individual privacy rights. 

\section{Ethical considerations } 
 This study was evaluated and approved as a human subject research by the Internal Review Board of the authors' university. 
Prior to starting the survey, participants were provided a link to the Internal Review Board document detailing:  the purpose of the study, the different procedures, any potential risks or discomforts that could occur during the study, potential benefits and compensation value and the mechanism of compensation delivery. We offered an incentive of \$7 per participant to complete our survey. The compensation was disbursed by Cint. 
Participants were also provided with an email and phone number to raise any concerns, or complaints related to the study. We did not receive any concerns or complains from the participants and there were no research-related injuries reported. 

The survey data collected was analyzed in an anonymous way. The names and emails of the participants were separated from their responses and each participant was assigned a Universally Unique Identifier (UUID) (which was generated randomly during the survey).
Names and emails of participants were only used to justify incentive expenses (compensation for participation) to our IRB board. 
When quoting participant responses from the free-form text replies in the survey, we identify each participant using the last 4 characters of their UUID. None of the Personally Identifiable Information (PII) will be made available publicly during or after the research is concluded.

\section{Compliance with Open science policy}
Literature has shown that data visualizations are better at conveying detailed information to participants when conducting perception surveys. To integrate interactive map visualizations in our survey, we developed a Flask app that we hosted on a private server during the experiment. Participants were recruited via Cint and then were taken to our server where the survey data was collected and consolidated. The Flask app is connected with a local database which has the questions and corresponding visualizations that are randomized and shown to participants. After the data collection, the analysis was performed using python. 

If required, we will be able to share: 1) \textbf{Hosting:} Flask app code, maria-db database with questions and visualizations as HTML files; 2) \textbf{Analysis} Python code/R code (regression analysis and Kruskal-Wallis test followed by Dunn post-hoc test). 
To enhance the reproducibility and replicability of our scientific findings, we can also share the anonymized responses to our survey. 

\bibliographystyle{unsrt}
\bibliography{arxiv}

\begin{thebibliography}{10}

\bibitem{walkingActivity}
Ruth~F Hunter, Leandro Garcia, Thiago~Herick de~Sa, Belen Zapata-Diomedi, Christopher Millett, James Woodcock, Alex~'sandy' Pentland, and Esteban Moro.
\newblock Effect of {COVID-19} response policies on walking behavior in {US} cities.
\newblock {\em Nat. Commun.}, 12(1):3652, June 2021.

\bibitem{hurricane_trd}
Jinpeng Wang and Yujie Hu.
\newblock Unraveling hurricane ian’s impact: A multiscale analysis of mobility networks in florida.
\newblock {\em Transportation Research Part D: Transport and Environment}, 136:104482, 2024.

\bibitem{homeworklocation}
Qingqing Chen and Ate Poorthuis.
\newblock Identifying home locations in human mobility data: an open-source r package for comparison and reproducibility.
\newblock {\em International Journal of Geographical Information Science}, 35(7):1425--1448, 2021.

\bibitem{home_work_poi_pipeline}
Katarzyna Siła-Nowicka, Jan Vandrol, Taylor Oshan, Jed~A. Long, Urška Demšar, and A.~Stewart Fotheringham.
\newblock Analysis of human mobility patterns from gps trajectories and contextual information.
\newblock {\em International Journal of Geographical Information Science}, 30(5):881--906, 2016.

\bibitem{FTCcase}
{F}{T}{C} settles unprecedented case against geolocation data broker --- therecord.media.
\newblock \url{https://therecord.media/ftc-settles-case-geolocation-data-broker-xmode-outlogic }.
\newblock [Accessed 27-08-2024].

\bibitem{brennen_loophole}
Brennan Center.
\newblock Closing the data broker loophole | brennan center for justice, Jan 2024.

\bibitem{grindr}
Sara Morrison.
\newblock This outed priest’s story is a warning for everyone about the need for data privacy laws, Jul 2021.

\bibitem{plannedparenthood}
Joseph Cox.
\newblock Data broker is selling location data of people who visit abortion clinics, May 2022.

\bibitem{colombia_privacy_study}
Margarita Gamarra, Inés Meriño~Fuentes, Juan Calabria~Sarmiento, Omar Gutierrez~Acosta, Mauricio Barrios~Barrios, Nallig Leal, and Pedro~Mario Wightman~Rojas.
\newblock Privacy perception in location-based services for mobile devices in the university community of the north coast of colombia.
\newblock {\em Ingenieria y Universidad}, 23(1), Feb 2019.

\bibitem{cross_platform_social_media_data}
Sarah Gilbert, Katie Shilton, and Jessica Vitak.
\newblock When research is the context: Cross-platform user expectations for social media data reuse.
\newblock {\em Big Data \& Society}, 10(1):20539517231164108, 2023.

\bibitem{martin}
Kirsten Martin and Helen Nissenbaum.
\newblock What is it about location?
\newblock {\em Berkeley technology law journal / Boalt Hall School of Law, University of California, Berkeley}, 12 2019.

\bibitem{safegraph}
SafeGraph.
\newblock Places data curated for accurate geospatial analytics, 2025.

\bibitem{spectusDeviceRecurring}
{D}evice {R}ecurring {A}reas and {S}ensitive {L}ocations - {S}pectus {D}ocumentation {P}ortal --- docs.spectus.ai.
\newblock \url{https://docs.spectus.ai/Getting%20Started/User_Guides/Data_Assets/Device_Recurring_Areas_and_Sensitive_Locations/#how-we-expose-the-devices-recurring-areas}.
\newblock [Accessed 05-02-2025].

\bibitem{longroadcomp}
Vincent Primault, Antoine Boutet, Sonia~Ben Mokhtar, and Lionel Brunie.
\newblock The long road to computational location privacy: A survey.
\newblock {\em IEEE Communications Surveys \& Tutorials}, 21(3):2772--2793, 2019.

\bibitem{ICADataCollection}
Jessica Vitak, Yuting Liao, Anouk Mols, Daniel Trottier, Michael Zimmer, Priya~C. Kumar, and Jason Pridmore.
\newblock When do data collection and use become a matter of concern? a cross-cultural comparison of u.
\newblock 2022.

\bibitem{nissenbaum2004privacy}
Helen Nissenbaum.
\newblock Privacy as contextual integrity.
\newblock {\em Wash. L. Rev.}, 79:119, 2004.

\bibitem{FTCsWarn}
{F}{T}{C}’s {K}han warns tech industry that agency will strictly enforce {A}{I} data privacy --- therecord.media.
\newblock \url{https://therecord.media/ftc-warns-tech-industry-ai-data-privacy}.
\newblock [Accessed 27-08-2024].

\bibitem{googleFusedLocation}
{F}used {L}ocation {P}rovider {A}{P}{I}  |  {G}oogle for {D}evelopers --- developers.google.com.
\newblock \url{https://developers.google.com/location-context/fused-location-provider}.
\newblock [Accessed 02-09-2024].

\bibitem{appleCCLocation}
Cclocationmanager | apple developer documentation.
\newblock \url{https://developer.apple.com/documentation/corelocation/cllocationmanager}.
\newblock [Accessed 02-09-2024].

\bibitem{vox_privacy}
Sara Morrison.
\newblock The hidden trackers in your phone, explained, Jul 2020.

\bibitem{contact_tracing_apps}
Alex Akinbi, Mark Forshaw, and Victoria Blinkhorn.
\newblock Contact tracing apps for the covid-19 pandemic: a systematic literature review of challenges and future directions for neo-liberal societies.
\newblock {\em Health Information Science and Systems}, 9(1), Apr 2021.

\bibitem{hong2016topic}
Lingzi Hong, Enrique Frias-Martinez, and Vanessa Frias-Martinez.
\newblock Topic models to infer socio-economic maps.
\newblock In {\em Proceedings of the AAAI Conference on Artificial Intelligence}, volume~30, 2016.

\bibitem{estabaan_urban_dynamics}
Yanni Yang, Alex Pentland, and Esteban Moro.
\newblock Identifying latent activity behaviors and lifestyles using mobility data to describe urban dynamics.
\newblock {\em EPJ Data Sci.}, 12(1):15, May 2023.

\bibitem{garcia_you_are_what_you_eat}
Bernardo Garcia-Bulle, Abigail~L. Horn, Brooke~M. Bell, Mohsen Bahrami, Burcin Bozkaya, Alex Pentland, Kayla de~la Haye, and Esteban Moro.
\newblock You are where you eat: Effect of mobile food environments on fast food visits.
\newblock {\em medRxiv}, 2022.

\bibitem{equalizer_social_infra}
Timothy Fraser, Takahiro Yabe, Daniel~P. Aldrich, and Esteban Moro.
\newblock The great equalizer? mixed effects of social infrastructure on diverse encounters in cities.
\newblock {\em Computers, Environment and Urban Systems}, 113:102173, 2024.

\bibitem{location_advertising}
Christine Bauer and Christine Strauss.
\newblock Location-based advertising on mobile devices.
\newblock {\em Management Review Quarterly}, 66(3):159–194, Jan 2016.

\bibitem{hong2019characterization}
Lingzi Hong, Jiahui Wu, Enrique Frias-Martinez, Andr{\'e}s Villarreal, and Vanessa Frias-Martinez.
\newblock Characterization of internal migrant behavior in the immediate post-migration period using cell phone traces.
\newblock In {\em Proceedings of the tenth international conference on information and communication technologies and development}, pages 1--12, 2019.

\bibitem{ma2014development}
Xiaolei Ma and Yinhai Wang.
\newblock Development of a data-driven platform for transit performance measures using smart card and gps data.
\newblock {\em Journal of Transportation Engineering}, 140(12):04014063, 2014.

\bibitem{wu2022enhancing}
Jiahui Wu, Saad~Mohammad Abrar, Naman Awasthi, Enrique Frias-Martinez, and Vanessa Frias-Martinez.
\newblock Enhancing short-term crime prediction with human mobility flows and deep learning architectures.
\newblock {\em EPJ data science}, 11(1):53, 2022.

\bibitem{ducham}
Roland Billen, Elsa Joao, and David Forrest, editors.
\newblock {\em Dynamic and mobile {GIS}}.
\newblock Innovations in GIS. CRC Press, Boca Raton, FL, November 2006.

\bibitem{employee_surveillance}
Jessica Vitak and Michael Zimmer.
\newblock Surveillance and the future of work: exploring employees' attitudes toward monitoring in a post-covid workplace.
\newblock {\em J. Comput. Mediat. Commun.}, 28, 2023.

\bibitem{mobile_data_thesis}
A~Gorra.
\newblock An analysis of the relationship between individuals? perceptions of privacy and mobile phone location data - a grounded theory study.
\newblock April 2007.

\bibitem{location_privacy_students}
Emily~Fekete Matthew~Haffner, Adam J.~Mathews and G.~Allen Finchum.
\newblock Location‐based social media behavior and perception: Views of university students.
\newblock {\em Geographical Review}, 108(2):203--224, 2018.

\bibitem{privacy_older_adults}
Rajarshi Chakraborty, Claire Vishik, and H.~Raghav Rao.
\newblock Privacy preserving actions of older adults on social media: Exploring the behavior of opting out of information sharing.
\newblock {\em Decision Support Systems}, 55(4):948–956, Nov 2013.

\bibitem{personality_location}
Iris Junglas and Christiane Spitzmuller.
\newblock Personality traits and privacy perceptions: An empirical study in the context of location-based services.
\newblock In {\em 2006 International Conference on Mobile Business}, pages 36--36, 2006.

\bibitem{personality_privacy_perception_2016}
Gaurav Bansal, Fatemeh~Mariam Zahedi, and David Gefen.
\newblock Do context and personality matter? trust and privacy concerns in disclosing private information online.
\newblock {\em Information \& Management}, 53(1):1--21, 2016.

\bibitem{location_privacy_SEM}
Edward Shih-Tse Wang and Ruenn-Lien Lin.
\newblock Perceived quality factors of location-based apps on trust, perceived privacy risk, and continuous usage intention.
\newblock {\em Behaviour \& Information Technology}, 36(1):2--10, 2017.

\bibitem{privacy_covid}
Baobao Zhang, Sarah Kreps, Nina McMurry, and R~Miles McCain.
\newblock Americans’ perceptions of privacy and surveillance in the covid-19 pandemic.
\newblock {\em Plos one}, 15(12):e0242652, 2020.

\bibitem{surveillance_covid}
Athina Ioannou and Iis Tussyadiah.
\newblock Privacy and surveillance attitudes during health crises: Acceptance of surveillance and privacy protection behaviours.
\newblock {\em Technology in Society}, 67:101774, 2021.

\bibitem{video_survellience}
Andrew Tzer-Yeu Chen, Morteza Biglari-Abhari, and Kevin I-Kai Wang.
\newblock Context is king: Privacy perceptions of camera-based surveillance.
\newblock In {\em 2018 15th IEEE International Conference on Advanced Video and Signal Based Surveillance (AVSS)}, pages 1--6, 2018.

\bibitem{location_advertisement}
Oliver~T. Kurtz, Bernd~W. Wirtz, and Paul~F. Langer.
\newblock An empirical analysis of location-based mobile advertising—determinants, success factors, and moderating effects.
\newblock {\em Journal of Interactive Marketing}, 54(1):69--85, 2021.

\bibitem{privacy_covid_2}
Junghwan Kim and Mei-Po Kwan.
\newblock An examination of people’s privacy concerns, perceptions of social benefits, and acceptance of covid-19 mitigation measures that harness location information: A comparative study of the us and south korea.
\newblock {\em ISPRS International Journal of Geo-Information}, 10(1):25, 2021.

\bibitem{incentive_participate_1}
Carlos Ochoa~Gómez.
\newblock Willingness to participate in geolocation-based research.
\newblock {\em PLOS ONE}, 17(12):e0278416, Dec 2022.

\bibitem{incentive_participate_2}
Florian Keusch, Bella Struminskaya, Christopher Antoun, Mick~P Couper, and Frauke Kreuter.
\newblock {Willingness to Participate in Passive Mobile Data Collection}.
\newblock {\em Public Opinion Quarterly}, 83(S1):210--235, 06 2019.

\bibitem{tradeoff}
Michael Benisch, Patrick~Gage Kelley, Norman Sadeh, and Lorrie~Faith Cranor.
\newblock Capturing location-privacy preferences: quantifying accuracy and user-burden tradeoffs.
\newblock {\em Personal Ubiquitous Comput.}, 15(7):679–694, October 2011.

\bibitem{privacy_exhaustion}
Vanessa Bracamonte, Sebastian Pape, and Sascha Loebner.
\newblock “all apps do this”: Comparing privacy concerns towards privacy tools and non-privacy tools for social media content.
\newblock {\em Proceedings on Privacy Enhancing Technologies}, 2022(3):57–78, Jul 2022.

\bibitem{privacy_risk_awareness}
Nina Gerber, Benjamin Reinheimer, and Melanie Volkamer.
\newblock Investigating people’s privacy risk perception.
\newblock {\em Proceedings on Privacy Enhancing Technologies}, 2019(3):267–288, Jul 2019.

\bibitem{privacy_prediction}
Yuchen Zhao, Juan Ye, and Tristan Henderson.
\newblock Privacy-aware location privacy preference recommendations.
\newblock In {\em Proceedings of the 11th International Conference on Mobile and Ubiquitous Systems: Computing, Networking and Services}, MOBIQUITOUS '14, page 120–129, Brussels, BEL, 2014. ICST (Institute for Computer Sciences, Social-Informatics and Telecommunications Engineering).

\bibitem{privacy_prediction2}
A.~Alshehri and Fayez Alotaibi.
\newblock Predicting users mobile app privacy preferences.
\newblock {\em Journal of Computer and Communications}, 2019.

\bibitem{privacy_prediction3}
Daniel Smullen, Yuanyuan Feng, Shikun Zhang, and N.~Sadeh.
\newblock The best of both worlds: Mitigating trade-offs between accuracy and user burden in capturing mobile app privacy preferences.
\newblock {\em Proceedings on Privacy Enhancing Technologies}, 2020:195 -- 215, 2020.

\bibitem{naeini}
Pardis~Emami Naeini, Sruti Bhagavatula, Hana Habib, Martin Degeling, Lujo Bauer, Lorrie~Faith Cranor, and Norman Sadeh.
\newblock Privacy expectations and preferences in an {IoT} world.
\newblock In {\em Thirteenth Symposium on Usable Privacy and Security (SOUPS 2017)}, pages 399--412, Santa Clara, CA, July 2017. USENIX Association.

\bibitem{contextualLabel}
Yaqing Yang, Tony~W Li, and Haojian Jin.
\newblock On the feasibility of predicting users' privacy concerns using contextual labels and personal preferences.
\newblock In {\em Proceedings of the CHI Conference on Human Factors in Computing Systems}, CHI '24, New York, NY, USA, 2024. Association for Computing Machinery.

\bibitem{PlacesofVisit}
Yanni Yang, Alex Pentland, and Esteban Moro.
\newblock Identifying latent activity behaviors and lifestyles using mobility data to describe urban dynamics.
\newblock {\em EPJ Data Sci.}, 12(1):15, May 2023.

\bibitem{poi_2}
Natalie Coleman, Chenyue Liu, Yiqing Zhao, and Ali Mostafavi.
\newblock Lifestyle pattern analysis unveils recovery trajectories of communities impacted by disasters.
\newblock {\em Humanit. Soc. Sci. Commun.}, 10(1), November 2023.

\bibitem{modesOfTransport_1}
Thomas Adler, Vince Bernardin, J.~Dumont, Leah Flake, and Hadi Sadrsadat.
\newblock The promise and limitations of locational app data for origin-destination analysis: A case study.
\newblock Technical Report FHWA-HEP-20-022, United States. Federal Highway Administration, 10 2017.

\bibitem{transport_NZ}
I-Ting Chuang, Lee Beattie, and Lei Feng.
\newblock Analysing the relationship between proximity to transit stations and local living patterns: A study of human mobility within a 15 min walking distance through mobile location data.
\newblock {\em Urban Science}, 7(4), 2023.

\bibitem{traj_analysis}
Wenwen Li, Shaohua Wang, Xiaoyi Zhang, Qingren Jia, and Yuanyuan Tian.
\newblock Understanding intra-urban human mobility through an exploratory spatiotemporal analysis of bike-sharing trajectories.
\newblock {\em International Journal of Geographical Information Science}, 34(12):2451--2474, 2020.

\bibitem{mode_detection_all}
Jinsoo Kim, Jae~Hun Kim, and Gunwoo Lee.
\newblock Gps data-based mobility mode inference model using long-term recurrent convolutional networks.
\newblock {\em Transportation Research Part C: Emerging Technologies}, 135:103523, 2022.

\bibitem{walk_nz}
I-Ting Chuang and Qingqing Chen.
\newblock Urban street dynamics: Assessing the relationship of sidewalk width and pedestrian activity in auckland, new zealand, based on mobile phone data.
\newblock {\em Urban Studies}, 0(0):00420980241293659, 0.

\bibitem{cuebiqInternational}
Cuebiq~Marketing Team.
\newblock {C}ome {B}ack {S}oon! {M}easuring the {E}ffectiveness of {T}ourism {C}ampaigns with --- cuebiq.com.
\newblock \url{https://cuebiq.com/come-back-soon-tourism/}.
\newblock [Accessed 06-02-2025].

\bibitem{vizbetter}
Florian~M. Farke, David~G. Balash, Maximilian Golla, Markus D{\"u}rmuth, and Adam~J. Aviv.
\newblock Are privacy dashboards good for end users? evaluating user perceptions and reactions to google{\textquoteright}s my activity.
\newblock In {\em 30th USENIX Security Symposium (USENIX Security 21)}, pages 483--500. USENIX Association, August 2021.

\bibitem{vignette_instructions}
Thom Baguley, Grace Dunham, and Oonagh Steer.
\newblock Statistical modelling of vignette data in psychology.
\newblock {\em British Journal of Psychology}, 113(4):1143--1163, 2022.

\bibitem{readingspeed}
Silvia Primativo, Donatella Spinelli, Pierluigi Zoccolotti, Maria De~Luca, and Marialuisa Martelli.
\newblock Perceptual and cognitive factors imposing ``speed limits'' on reading rate: A study with the rapid serial visual presentation.
\newblock {\em PLoS One}, 11(4):e0153786, April 2016.

\bibitem{census_education_attainment}
{U.S. Census Bureau}.
\newblock Age and sex.
\newblock American Community Survey 5-Year Estimates Subject Tables, Table S0101, 2022.
\newblock Available at: \url{https://data.census.gov/table/ACSST1Y2023.S1501?q=education%20attainment}.

\bibitem{CISmartHome}
Noah Apthorpe, Yan Shvartzshnaider, Arunesh Mathur, Dillon Reisman, and Nick Feamster.
\newblock Discovering smart home internet of things privacy norms using contextual integrity.
\newblock {\em Proc. ACM Interact. Mob. Wearable Ubiquitous Technol.}, 2(2), jul 2018.

\bibitem{carpenterEvade}
Dori~H Rahbar.
\newblock How the government's purchase of commercial location data violates carpenter and evades the fourth amendment.
\newblock {\em Columbia Law Rev.}, 122(3):713--754, 2022.

\bibitem{IUIPC}
Naresh~K Malhotra, Sung~S Kim, and James Agarwal.
\newblock Internet users' information privacy concerns ({IUIPC)}: The construct, the scale, and a causal model.
\newblock {\em Inf. Syst. Res.}, 15(4):336--355, December 2004.

\end{thebibliography}

\appendix

\begin{table}[ht]
\centering
\caption{Age distribution in the U.S. and in our survey population.}
\begin{tabular}{lll}
\toprule
Age   & Census \% & Survey \% \\
\midrule
18-24 & 12\%           & 8\%       \\
25-34 & 18\%           & 22\%      \\
35-44 & 17\%           & 19\%      \\
45-54 & 16\%           & 17\%      \\
55-64 & 17\%           & 19\%      \\
65+   & 21\%           & 14\%     \\
\bottomrule
\end{tabular}

\label{tab:agedistribution}
\end{table}

\begin{table}[ht]
\centering
\caption{Gender distribution for the U.S. Adult population (age 18 and above) and in our survey population.}
\begin{tabular}{lll}
\toprule
Gender & U.S. Census & Survey \\
\midrule
Male   & 49\%      & 46\% \\
Female & 51\%      & 53\% \\ 
\bottomrule
\end{tabular}
\label{tab:genderdistribution}
\end{table}

\begin{table*}[]
\centering
\caption{Percentage of responses for each privacy attitude question across levels of comfort. }
\begin{tabular}{|p{0.18\textwidth}p{0.10\textwidth}|p{0.10\textwidth}p{0.10\textwidth}p{0.06\textwidth}p{0.09\textwidth}p{0.09\textwidth}|}

\multicolumn{2}{|c|}\textbf{Privacy Attitudes}  & \multicolumn{5}{c|}{\textbf{Comfort Level}} \\
    \cline{3-7}
    & & \textbf{Very Uncomfortable} & \textbf{Uncomfortable} & \textbf{Neutral} & \textbf{Comfortable} & \textbf{Very Comfortable} \\
\cline{1-7}
\multirow{5}{*}{\textbf{I trust businesses}} &  \textbf{S. disagree}        & \cellcolor[HTML]{72C6E9}37.872 & \cellcolor[HTML]{BDE4F5}17.872 & \cellcolor[HTML]{D3EDF8}11.915 & \cellcolor[HTML]{C9EAF7}14.468 & \cellcolor[HTML]{BDE4F5}17.872 \\
& \textbf{Disagree}                 & \cellcolor[HTML]{A7DCF1}23.62  & \cellcolor[HTML]{97D5EF}27.964 & \cellcolor[HTML]{B6E2F4}19.548 & \cellcolor[HTML]{B5E1F4}19.91  & \cellcolor[HTML]{DEF2FA}8.959  \\
& \textbf{Undecided}                & \cellcolor[HTML]{CDEBF7}13.6   & \cellcolor[HTML]{AEDEF2}21.76  & \cellcolor[HTML]{88CFEC}31.893 & \cellcolor[HTML]{A6DBF1}24.053 & \cellcolor[HTML]{DFF2FA}8.693  \\
& \textbf{Agree}                    & \cellcolor[HTML]{DAF0F9}10.047 & \cellcolor[HTML]{CAEAF7}14.317 & \cellcolor[HTML]{B0DFF3}21.193 & \cellcolor[HTML]{6FC5E8}38.556 & \cellcolor[HTML]{C4E7F6}15.887 \\
& \textbf{S. Agree}         & \cellcolor[HTML]{DFF2FA}8.73   & \cellcolor[HTML]{E9F6FC}6.032  & \cellcolor[HTML]{E0F3FA}8.413  & \cellcolor[HTML]{96D5EF}28.095 & \cellcolor[HTML]{49B5E2}48.73  \\
\cline{1-7}
\multirow{5}{*}{\textbf{Privacy   importance}} & \textbf{S. disagree}        & \cellcolor[HTML]{AADDF2}22.857 & \cellcolor[HTML]{C5E8F6}15.714 & \cellcolor[HTML]{E5F5FB}7.143  & \cellcolor[HTML]{C5E8F6}15.714 & \cellcolor[HTML]{6FC5E8}38.571 \\
& \textbf{Disagree}                 & \cellcolor[HTML]{EDF8FC}5      & \cellcolor[HTML]{CFECF8}13     & \cellcolor[HTML]{A9DDF2}23     & \cellcolor[HTML]{84CDEC}33     & \cellcolor[HTML]{9ED8F0}26     \\
& \textbf{Undecided}                & \cellcolor[HTML]{F1FAFD}3.939  & \cellcolor[HTML]{E8F6FC}6.364  & \cellcolor[HTML]{86CEEC}32.424 & \cellcolor[HTML]{75C7E9}36.97  & \cellcolor[HTML]{B4E1F3}20.303 \\
& \textbf{Agree}                    & \cellcolor[HTML]{E5F5FB}7.173  & \cellcolor[HTML]{C3E7F6}16.14  & \cellcolor[HTML]{A4DAF1}24.522 & \cellcolor[HTML]{6DC4E8}39.103 & \cellcolor[HTML]{CFECF8}13.06  \\
& \textbf{S. Agree}         & \cellcolor[HTML]{B8E2F4}19.218 & \cellcolor[HTML]{B5E1F3}20.025 & \cellcolor[HTML]{B3E1F3}20.328 & \cellcolor[HTML]{A7DBF1}23.733 & \cellcolor[HTML]{C1E6F5}16.696 \\
\cline{1-7}
\multirow{5}{*}{\textbf{Trust   in government}}& \textbf{S. disagree}        & \cellcolor[HTML]{84CDEC}32.898 & \cellcolor[HTML]{A4DAF1}24.408 & \cellcolor[HTML]{BEE5F5}17.551 & \cellcolor[HTML]{C0E6F5}17.061 & \cellcolor[HTML]{E1F3FB}8.082  \\
& \textbf{Disagree}                 & \cellcolor[HTML]{C9E9F7}14.579 & \cellcolor[HTML]{A3DAF1}24.735 & \cellcolor[HTML]{A5DBF1}24.112 & \cellcolor[HTML]{9ED8F0}26.168 & \cellcolor[HTML]{D9F0F9}10.405 \\
& \textbf{Undecided}                & \cellcolor[HTML]{D5EEF9}11.299 & \cellcolor[HTML]{C5E8F6}15.71  & \cellcolor[HTML]{93D3EE}29.124 & \cellcolor[HTML]{8CD0ED}30.937 & \cellcolor[HTML]{CFECF8}12.931 \\
& \textbf{Agree}                    & \cellcolor[HTML]{E8F6FC}6.205  & \cellcolor[HTML]{CEEBF8}13.179 & \cellcolor[HTML]{B1E0F3}20.872 & \cellcolor[HTML]{6BC3E8}39.744 & \cellcolor[HTML]{B5E1F3}20     \\
& \textbf{S. Agree}         & \cellcolor[HTML]{EAF7FC}5.882  & \cellcolor[HTML]{E6F5FB}6.723  & \cellcolor[HTML]{CEEBF8}13.277 & \cellcolor[HTML]{86CEEC}32.605 & \cellcolor[HTML]{64C0E7}41.513 \\
\cline{1-7}
\multirow{5}{*}{\makecell{\textbf{Great that young}\\\textbf{people are prepared to}\\\textbf{defy authority}}}& \textbf{S. disagree}
& \cellcolor[HTML]{C0E6F5}16.978 & \cellcolor[HTML]{CBEAF7}14.044 & \cellcolor[HTML]{B6E2F4}19.733 & \cellcolor[HTML]{A5DBF1}24.178 & \cellcolor[HTML]{A2D9F0}25.067 \\
& \textbf{Disagree}                 & \cellcolor[HTML]{CFECF8}13.048 & \cellcolor[HTML]{B9E3F4}18.974 & \cellcolor[HTML]{AADDF2}22.735 & \cellcolor[HTML]{7ECBEB}34.644 & \cellcolor[HTML]{D8EFF9}10.598 \\
& \textbf{Undecided}                & \cellcolor[HTML]{CCEBF7}13.69  & \cellcolor[HTML]{B3E0F3}20.428 & \cellcolor[HTML]{A3DAF1}24.599 & \cellcolor[HTML]{92D3EE}29.198 & \cellcolor[HTML]{D2EDF8}12.086 \\
& \textbf{Agree}                    & \cellcolor[HTML]{DAF0FA}9.969  & \cellcolor[HTML]{BAE3F4}18.7   & \cellcolor[HTML]{A9DCF2}23.034 & \cellcolor[HTML]{81CCEB}33.746 & \cellcolor[HTML]{C9E9F7}14.551 \\
& \textbf{S. Agree}         & \cellcolor[HTML]{AFDFF3}21.504 & \cellcolor[HTML]{D4EEF8}11.729 & \cellcolor[HTML]{BEE5F5}17.594 & \cellcolor[HTML]{B1E0F3}20.902 & \cellcolor[HTML]{96D5EF}28.271 \\
\cline{1-7}
\multirow{5}{*}{\makecell{\textbf{Discipline and}\\\textbf{follow leader}}} & \textbf{S. disagree}        & \cellcolor[HTML]{88CFEC}32.048 & \cellcolor[HTML]{BAE3F4}18.554 & \cellcolor[HTML]{B8E3F4}19.036 & \cellcolor[HTML]{B8E3F4}19.036 & \cellcolor[HTML]{D5EEF9}11.325 \\
& \textbf{Disagree}                 & \cellcolor[HTML]{C5E8F6}15.767 & \cellcolor[HTML]{A8DCF2}23.313 & \cellcolor[HTML]{B4E1F3}20.307 & \cellcolor[HTML]{93D3EE}29.08  & \cellcolor[HTML]{D4EEF9}11.534 \\
& \textbf{Undecided}                & \cellcolor[HTML]{CFECF8}13.023 & \cellcolor[HTML]{BBE3F4}18.43  & \cellcolor[HTML]{93D3EE}29.07  & \cellcolor[HTML]{96D4EF}28.314 & \cellcolor[HTML]{D6EFF9}11.163 \\
& \textbf{Agree}                    & \cellcolor[HTML]{E2F4FB}7.795  & \cellcolor[HTML]{C3E7F6}16.256 & \cellcolor[HTML]{A9DCF2}23.128 & \cellcolor[HTML]{75C7E9}37.026 & \cellcolor[HTML]{C4E7F6}15.795 \\
& \textbf{S. Agree}         & \cellcolor[HTML]{DEF2FA}9      & \cellcolor[HTML]{DCF1FA}9.444  & \cellcolor[HTML]{C9EAF7}14.444 & \cellcolor[HTML]{90D2EE}29.889 & \cellcolor[HTML]{74C7E9}37.222 \\
\cline{1-7}
\multirow{5}{*}{\textbf{Give benefit of doubt}}& \textbf{S. disagree}        & \cellcolor[HTML]{6AC3E7}40     & \cellcolor[HTML]{BEE5F5}17.6   & \cellcolor[HTML]{C4E7F6}16     & \cellcolor[HTML]{D6EEF9}11.2   & \cellcolor[HTML]{C7E8F6}15.2   \\
& \textbf{Disagree}                 & \cellcolor[HTML]{B6E1F4}19.744 & \cellcolor[HTML]{AADDF2}22.821 & \cellcolor[HTML]{B9E3F4}18.974 & \cellcolor[HTML]{B1E0F3}21.026 & \cellcolor[HTML]{BEE5F5}17.436 \\
& \textbf{Undecided}                & \cellcolor[HTML]{BFE5F5}17.253 & \cellcolor[HTML]{BEE5F5}17.582 & \cellcolor[HTML]{95D4EE}28.352 & \cellcolor[HTML]{95D4EE}28.462 & \cellcolor[HTML]{E0F3FA}8.352  \\
& \textbf{Agree}                    & \cellcolor[HTML]{D0ECF8}12.637 & \cellcolor[HTML]{B7E2F4}19.264 & \cellcolor[HTML]{AADDF2}22.969 & \cellcolor[HTML]{85CEEC}32.755 & \cellcolor[HTML]{D1EDF8}12.375 \\
& \textbf{S. Agree}         & \cellcolor[HTML]{D4EEF8}11.756 & \cellcolor[HTML]{D2EDF8}12.258 & \cellcolor[HTML]{BCE4F5}17.993 & \cellcolor[HTML]{9BD7EF}26.953 & \cellcolor[HTML]{8BD0ED}31.039 \\
\cline{1-7}
\end{tabular}
\label{tab:attitudes_responses}
\end{table*}

\begin{table}[h]
    \centering
    \caption{Actor and Comfortableness (Mean)}
    \begin{tabular}{lr}
        \toprule
        Actor & Answer \\
        \midrule
        Actor: Commercial Entity & 0.162 \\
        Actor: Federal government agency & -0.073 \\
        Actor: Law enforcement agency & 0.081 \\
        Actor: Local government agency & 0.189 \\
        Actor: Academic researchers & 0.387 \\
        Actor: Emergency services & 0.525 \\
        Actor: Doctor & 0.256 \\
        Actor: Employer & -0.424 \\
        Actor: Family & 0.383 \\
        \bottomrule
    \end{tabular}
    \label{tab:actorcomfortmean}
\end{table}
\begin{table}[h]
\centering
\caption{Purpose and Comfortableness (Means)}
\begin{tabular}{p{0.70\linewidth}p{0.2\linewidth}}
\toprule
Purpose &  Answer \\
\midrule
Purpose: Analysis of terrorist attacks &        0.126 \\
Purpose: Analysis of Public transit services &        0.276 \\
Purpose: Control spread of diseases &        0.233 \\
Purpose: Monitor mobility patterns &       -0.022 \\
Purpose: Personal wellness &        0.157 \\
Purpose: Show ads &        0.185 \\
Purpose: Analysis of criminal activity &        0.147 \\
Purpose: Design new walking/cycling infrastructure &        0.425 \\
Purpose: Identify locations for infrastructure &        0.446 \\
Purpose: Optimize work productivity &       -0.054 \\
\bottomrule
\end{tabular}
    \label{tab:purposecomfortmean}
\end{table}

\begin{table}[h]
\centering
\caption{Feature and Comfortableness (Means)}
\begin{tabular}{p{0.70\linewidth}p{0.2\linewidth}}
\toprule
Feature &  Answer \\
\midrule
Feature: International visits (Detailed) &        0.286 \\
Feature: International visits (Obfuscated) &        0.247 \\
\makecell[l]{Feature: Area you spent most \\  of your time (Obfuscated)} &        0.200 \\
Feature: Modes of transportation (Obfuscated) &        0.220 \\
Feature: Modes of transportation (Detailed) &        0.175 \\
Feature: Places you visit (Detailed) &        0.092 \\
Feature: Places you visit (Obfuscated) &        0.224 \\
Feature: Home location (Obfuscated) &        0.210 \\
Feature: Home location (Detailed) &        0.155 \\
Feature: Work location (Obfuscated) &        0.263 \\
Feature: Work location (Detailed) &        0.268 \\
Feature: Least frequent trips (Detailed) &        0.026 \\
\makecell[l]{Feature: Most frequent trips between \\ counties (Obfuscated)} &        0.129 \\
\makecell[l]{Feature: Most frequent trips between \\ counties('Google') (Obfuscated)} &        0.152 \\
Feature: Most frequent trips (Detailed) &       -0.058 \\
\makecell[l]{Feature: Most frequent type \\ of trips (Obfuscated)} &        0.145 \\
Feature: Frequent walking activity (Detailed) &        0.054 \\
Feature: Frequent walking activity (Obfuscated) &        0.356 \\
\bottomrule
\end{tabular}
\label{tab:featcomfortmean}
\end{table}

\begin{figure}[ht]
    \centering
\includegraphics[width=0.5\textwidth]{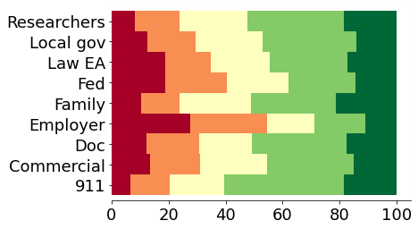}
    \caption{Percentage of responses per actor and level of comfort (ordered left to right: Very uncomfortable, Uncomfortable, Neutral, Comfortable, Very comfortable). }
    \label{fig:actor_responses}
\end{figure}

\begin{figure}[ht]
    \centering
\includegraphics[width=0.5\textwidth]{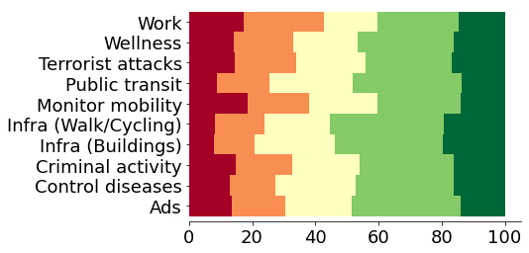}
    \caption{Percentage of responses per purpose and level (ordered left to right: Very uncomfortable, Uncomfortable, Neutral, Comfortable, Very comfortable). }
    \label{fig:purpose_responses}
\end{figure}

\section{Ordinal Regression}
\label{section:ordinal_regression}
The survey measures comfort levels using a five-point ordinal scale (very uncomfortable $\leq$ uncomfortable $\leq$ neutral $\leq$ comfortable $\leq$ very comfortable). Mixed effects Ordinal regression was chosen over ordinal regression and multi-class classification to account for the ordered nature of the dependent variable.
Ordinal regressions are preferred to linear regressions because ordinal categories are not necessarily evenly spaced (e.g., the difference between uncomfortable and neutral may not equal the difference between neutral and comfortable).

We used the following formula for the Mixed effects Ordinal regression: 


 \begin{align}
 \label{eq:regression}
 \text{Ans} &\sim 1 + \text{Actor} + \text{Purpose} + \text{Feature} + \notag \\
&\text{Ethnicity} + \text{Education} + \notag \\
&\text{Privacy and Attitude questions} + \notag   +\\
& (1|\text{ID}) + (1|\text{apf})
\end{align}

\textbf{Assumption Testing.} We test the assumptions for the mixed effects ordinal regression model: proportionality odds and multicollinearity using the nominal\_test and scale\_test in R. None of the variables have p-value < 0.05 (thus no significant evidence of non-proportional odds). We also compute Variance Inflation Factor (VIF) and all the variables have VIF < 2 which indicates no multicollinearity in the regression setting.

Next, we also check goodness of fit and compare different ordinal regression settings in Table \ref{tab:model_aic_compare}. We find that the model with best AIC is the one with two random intercepts ID and vignette (actor-purpose-feature combination). ID is the UUID of an individual (as we ask multiple questions to the same person). Actor-Purpose-Feature (1|apf) models the between vignette heterogeneity as described in \cite{vignette_instructions}. We analyze this regression model in the paper.

\textbf{Weights.} We weigh the samples in the mixed effects ordinal regression model for representativity of education and ethnicity. We calculate these weights by dividing \% of expected population by our population e.g.,: 6.2\% of adult white population is educated and has a degree under high school. Since we have 3.6\% participants in that strata, we assign a weight of 1.72 to each respondent. 

\begin{table}[ht]
\centering
\caption{ANOVA analysis of different models to select the best model. CLM: Cumulative link model (Ordinal model), CLMM (cumulative linked mixed model). Best models are $CLMM_{ID}$ and $CLMM_{ID\_VIG}$ with lowest AIC scores. Significance "**" < 0.05}
\begin{tabularx}{0.5\textwidth}{lXlll} 
Model & Description  & AIC \\
\hline
CLM & a+p+f+privacy and attitudes & 20218  \\
\hline
$CLMM_{intercept}$ & random intercepts only & 19332 \\
\hline
$CLMM_{noapf}$ & Random Intercept + privacy \& attitudes & 18803** \\
\hline
$CLMM_{ID}$ & a+p+f+privacy and attitudes + (1|ID)  & \textbf{18589**} \\
\hline
$CLMM_{ID\_VIG}$ & a+p+f+privacy and attitudes + (1|ID)+ (1|apf)  & \textbf{18583**} 
\end{tabularx}
\label{tab:model_aic_compare}
\end{table}

\begin{table*}[ht]
\centering
\footnotesize
\caption{Additional control variables used in the regression setting which were significant (** pvalue<0.05). L :Linear, C : Cubic, Q: Quadratic. }
\begin{tabular}{llllll}
& Coefficients & Odds Ratio & Std Err & Statistic & p val \\
Privacy.importance.L                                                            & \textbf{-1.301** }    & 0.272      & 0.385   & -3.382    & 0.001 \\
Privacy.importance.Q                                                            & -0.246       & 0.782      & 0.353   & -0.697    & 0.486 \\
Privacy.importance.C                                                            & 0.141        & 1.151      & 0.336   & 0.42      & 0.675 \\
Privacy.importance\textasciicircum{}4                                           & -0.048       & 0.953      & 0.285   & -0.168    & 0.867 \\
Trust.in.government.L                                                           & \textbf{1.398** }     & 4.047      & 0.189   & 7.386     & 0     \\
Trust.in.government.Q                                                           & -0.172       & 0.842      & 0.158   & -1.087    & 0.277 \\
Trust.in.government.C                                                           & 0.134        & 1.143      & 0.122   & 1.105     & 0.269 \\
Trust.in.government\textasciicircum{}4                                          & -0.064       & 0.938      & 0.109   & -0.588    & 0.557 \\
Great.that.young.people.are.prepared.to.defy.authority.L                        & \textbf{-0.425**}     & 0.654      & 0.161   & -2.641    & 0.008 \\
Great.that.young.people.are.prepared.to.defy.authority.Q                        & 0.143        & 1.154      & 0.149   & 0.955     & 0.339 \\
Great.that.young.people.are.prepared.to.defy.authority.C                        & \textbf{-0.38**}      & 0.684      & 0.119   & -3.191    & 0.001 \\
Great.that.young.people.are.prepared.to.defy.authority\textasciicircum{}4       & -0.033       & 0.968      & 0.108   & -0.304    & 0.761 \\
Discipline.and.follow.leader.L                                                  & \textbf{0.753**}      & 2.123      & 0.168   & 4.493     & 0     \\
Discipline.and.follow.leader.Q                                                  & 0.097        & 1.102      & 0.15    & 0.648     & 0.517 \\
Discipline.and.follow.leader.C                                                  & \textbf{0.26**}       & 1.297      & 0.119   & 2.188     & 0.029 \\
Discipline.and.follow.leader\textasciicircum{}4                                 & -0.115       & 0.891      & 0.108   & -1.058    & 0.29  \\
Give.benefit.of.doubt.L                                                         & 0.475        & 1.608      & 0.288   & 1.651     & 0.099 \\
Give.benefit.of.doubt.Q                                                         & 0.075        & 1.078      & 0.253   & 0.297     & 0.767 \\
Give.benefit.of.doubt.C                                                         & \textbf{0.433**}      & 1.542      & 0.201   & 2.158     & 0.031 \\
Give.benefit.of.doubt\textasciicircum{}4                                        & \textbf{-0.516**}     & 0.597      & 0.165   & -3.131    & 0.002 \\
Avoid.mobile.apps.due.to.privacy.concern.L                                      & \textbf{-0.783**}     & 0.457      & 0.136   & -5.777    & 0     \\
Avoid.mobile.apps.due.to.privacy.concern.Q                                      & -0.085       & 0.919      & 0.097   & -0.879    & 0.379 \\
Uninstalled.app.for.collecting.data.L                                           & -0.159       & 0.853      & 0.133   & -1.192    & 0.233 \\
Uninstalled.app.for.collecting.data.Q                                           & 0.01         & 1.01       & 0.096   & 0.105     & 0.917 \\
Computer.Science.or.Programming.Experience.L                                    & \textbf{0.558**}      & 1.747      & 0.24    & 2.321     & 0.02  \\
Computer.Science.or.Programming.Experience.Q                                    & 0.058        & 1.06       & 0.163   & 0.355     & 0.723 \\
Computer.Science.or.Programming.Experience.C                                    & 0.042        & 1.043      & 0.114   & 0.367     & 0.714 \\
Machine.Learning.Experience.L                                                   & 0.019        & 1.019      & 0.289   & 0.067     & 0.947 \\
Machine.Learning.Experience.Q                                                   & 0.057        & 1.059      & 0.194   & 0.296     & 0.767 \\
Machine.Learning.Experience.C                                                   & -0.015       & 0.985      & 0.134   & -0.114    & 0.909 \\
Statistics.Experience.L                                                         & 0.05         & 1.051      & 0.276   & 0.18      & 0.857 \\
Statistics.Experience.Q                                                         & -0.134       & 0.875      & 0.191   & -0.7      & 0.484 \\
Statistics.Experience.C                                                         & -0.13        & 0.878      & 0.127   & -1.022    & 0.307 \\
Frequency.of.use.of.data.collection.tools.like.GMaps..Fitbit.L                  & 0.321        & 1.379      & 0.167   & 1.926     & 0.054 \\
Frequency.of.use.of.data.collection.tools.like.GMaps..Fitbit.Q                  & 0.212        & 1.236      & 0.162   & 1.305     & 0.192 \\
Frequency.of.use.of.data.collection.tools.like.GMaps..Fitbit.C                  & -0.079       & 0.924      & 0.125   & -0.627    & 0.531 \\
Frequency.of.use.of.data.collection.tools.like.GMaps..Fitbit\textasciicircum{}4 & -0.195       & 0.823      & 0.14    & -1.393    & 0.163 \\
Privacy.settings.awareness.L                                                    & \textbf{0.461** }     & 1.586      & 0.179   & 2.571     & 0.01  \\
Privacy.settings.awareness.Q                                                    & 0.076        & 1.079      & 0.156   & 0.484     & 0.628 \\
Privacy.settings.awareness.C                                                    & -0.019       & 0.981      & 0.117   & -0.166    & 0.868 \\
Privacy.settings.awareness\textasciicircum{}4                                   & -0.133       & 0.875      & 0.12    & -1.104    & 0.269 \\
I.trust.businessess.L                                                           & \textbf{1.159**}      & 3.187      & 0.259   & 4.484     & 0     \\
I.trust.businessess.Q                                                           & \textbf{0.62**}       & 1.859      & 0.222   & 2.791     & 0.005 \\
I.trust.businessess.C                                                           & 0.163        & 1.177      & 0.153   & 1.068     & 0.286 \\
I.trust.businessess\textasciicircum{}4                                          & 0.085        & 1.089      & 0.115   & 0.741     & 0.459
\end{tabular}
\label{tab:attitude_privacy_coeffs}
\end{table*}


\section{Kruskal Wallis and Dunn posthoc test}
\label{section:KWDunn}
The Kruskal-Wallis test, a non-parametric method, is used to identify statistically significant differences in medians across multiple independent groups. We use the "survey" package in R to factor for the random intercepts while calculating the Kruskal-wallis statistic. While this test determines the presence of differences, it does not specify which pairs differ. Therefore, we conduct a post-hoc Dunn test to identify pairs of groups with statistically significant median differences. This combined approach is employed throughout our statistical analysis to discern perception differences. We do not use interaction effects in regression settings because in interactions, the significant coefficients indicate odds ratio compared to only the baseline interaction e.g., actor:education interactions that are significant in the regression setting would only be significantly higher/lower from the baseline (Commercial Entity: Highschool to Bachelors). The comparisons for all actors:education pairs cannot be established with a single regression model. KW with Dunn can do pairwise comparisons and post hoc corrections to reduce FDR (False Discovery Rate). 

We demonstrate the process using an example. Let us assume we are interested in understanding significant differences in comfort levels across actors and location features. 
Participants were asked about their comfort levels in sharing location data for different actors and features. Each actor-feature combination (N=138) represents an independent group, allowing for comparison using the Kruskal-Wallis test. The test yielded an F statistic of [316.28] with p-value < 0.05, indicating significant differences between the medians of actor-feature pairs.
Following the Kruskal-Wallis test, we employed the Dunn post-hoc test with the Benjamini-Hochberg (to reduce false discovery rate) correction for multiple comparisons. This post-hoc analysis produces an N x N matrix (where N is the number of actor-feature combinations), with each cell containing the adjusted p-value for the corresponding pairwise comparison.
For each actor, we tabulate only those features whose median comfort levels were found to be \textbf{statistically significantly different (pvalue < 0.05) by at least 1}. By grouping the results by actors and analyzing the different features per actor, we can understand how perceptions of the same actor may vary when accessing different types of location data.
This analytical approach is applied throughout the paper for various combinations, including actor-purpose, purpose-feature, education-feature, and privacy-actor pairs, to provide a comprehensive understanding of the factors influencing location data sharing comfort levels.

\begin{table}[]
\centering
\footnotesize
\caption{KW with Dunn test to identify significantly different medians when different actors and purposes are involved with the same feature. Median differences are statistically different (pval <0.01)
Kruskal-Wallis Chi-sq statistic interaction(feature,actor)= 316.28 pval<0.01; interaction(feature,purpose)= 236.13 pval<0.01;
}
\begin{tabular}{|p{0.18\linewidth}|p{0.17\linewidth}|p{0.18\linewidth}|p{0.06\linewidth}|p{0.05\linewidth}|p{0.08\linewidth}|}
\hline
Feature & Actor1   & Actor2   & M1 & M2 & \text{\footnotesize{M2-M1}} \\ \hline
Radius(Ob)                 & Employer                   & Researchers                & -1                           & 1                            & 2                               \\ \cline{3-6}
&                               & Commercial                 & -1                           & 0                            & 1                               \\ \cline{3-6}
&                               & 911                        & -1                           & 1                            & 2                               \\ \cline{3-6}
&                               & Family                     & -1                           & 1                            & 2                               \\ \cline{3-6}
&                               & Local gov                  & -1                           & 1                            & 2                               \\ \cline{1-6}
Radius(Ob)                 & Fed                        & 911                        & 0                            & 1                            & 1                               \\ \cline{1-6}
Freq Walks(Ob)             & Employer                   & Researchers                & 0                            & 1                            & 1                               \\ \cline{3-6}
&                               & Doc                        & 0                            & 1                            & 1                               \\ \cline{3-6}
&                               & Family                     & 0                            & 1                            & 1                               \\ \cline{1-6}
Freq Walks(Ob)             & Fed                        & Researchers                & 0                            & 1                            & 1                               \\ \cline{3-6}
&                               & Doc                        & 0                            & 1                            & 1                               \\ \cline{3-6}
&                               & Family                     & 0                            & 1                            & 1                               \\ \cline{1-6}
Freq Walks(Ob)             & Local gov                  & Doc                        & 0                            & 1                            & 1                               \\ \cline{3-6}
Home(D)                    & Commercial                 & Local gov                  & 0                            & 1                            & 1                               \\ \cline{3-6}
& Fed                        & Local gov                  & -0.5                         & 1                            & 1.5                             \\ \cline{1-6}
Least FreqT(D)             & Employer                   & Researchers                & -1                           & 1                            & 2                               \\ \cline{3-6}
&                               & Commercial                 & -1                           & 0                            & 1                               \\ \cline{3-6}
&                               & 911                        & -1                           & 1                            & 2                               \\ \cline{1-6}
Transport-    & Employer                   & Researchers                & 0                            & 1                            & 1                               \\
ation(Ob) & & & & & \\
\cline{1-6}
FreqT(D)                   & Employer                   & Researchers                & -1                           & 0                            & 1                               \\ \cline{3-6}
&                               & 911                        & -1                           & 1                            & 2                               \\ \cline{1-6}
FreqT-counties(Ob)         & Employer                   & Researchers                & -1                           & 1                            & 2                               \\ \cline{3-6}
&                               & Commercial                 & -1                           & 0                            & 1                               \\ \cline{3-6}
&                               & 911                        & -1                           & 1                            & 2                               \\ \cline{3-6}
&                               & Law EA                     & -1                           & 0                            & 1                               \\ \cline{3-6}
&                               & Local gov                  & -1                           & 1                            & 2                               \\ \cline{1-6}
FreqT-counties(Ob)         & Fed                        & Researchers                & -0.5                         & 1                            & 1.5                             \\ \cline{3-6}
&                               & 911                        & -0.5                         & 1                            & 1.5                             \\ \cline{3-6}
&                               & Local gov                  & -0.5                         & 1                            & 1.5                             \\ \cline{1-6}
FreqTypeT(D)               & Employer                   & Researchers                & -0.5                         & 1                            & 1.5                             \\ \cline{1-6}
Visits(D)                  & Employer                   & Researchers                & -1                           & 1                            & 2                               \\ \cline{3-6}
&                               & Commercial                 & -1                           & 0                            & 1                               \\ \cline{3-6}
&                               & Doc                        & -1                           & 1                            & 2                               \\  \cline{3-6}
&                               & 911                        & -1                           & 1                            & 2                               \\  \cline{1-6}
Visits(Ob)                 & Employer                   & Researchers                & -1                           & 1                            & 2                               \\ \cline{3-6}
&                               & Commercial                 & -1                           & 0                            & 1                               \\ \cline{3-6}
&                                & Family                     & -1                           & 1                            & 2                               \\\cline{3-6}
&                               & Law EA                     & -1                           & 0                            & 1                               \\ \cline{3-6}
&                               & Local gov                  & -1                           & 0                            & 1                               \\ \cline{1-6}
Feature & Purpose1   & Purpose2   & M1 & M2 & \text{\footnotesize{M2-M1}} \\ \cline{1-6}
Visits(Ob)                 & Monitor mobility           & Public transit             & 0                            & 1                            & 1                               \\ \cline{3-6}
&                               & Infra (Walk/Cycling)       & 0                            & 1                            & 1               \\
\cline{1-6}
\end{tabular}
\label{tab:kw_featureXactor_featureXpurpose_BIG}
\end{table}

\begin{table}[]
\centering
\footnotesize
\caption{KW with Dunn test to identify significantly different medians when different features are involved with an actor. Median differences are statistically different (pvalue <0.01). Kruskal wallis chi-sq interaction(feature,actor)= 316.28 pval<0.01
}
\begin{tabularx}{\linewidth}{|p{0.15\linewidth}|p{0.20\linewidth}|p{0.23\linewidth}|p{0.06\linewidth}|p{0.03\linewidth}|p{0.07\linewidth}|}
 \cline{1-6}
Actors        & Feature1              & Feature2            & M1 & M2 & \text{\footnotesize{M2-M1}} \\  \cline{1-6}
Commercial & FreqT(D)           & Freq Walks(Ob)   & 0       & 1       & 1          \\ \cline{1-6}
Doc        & Radius(Ob)         & Freq Walks(Ob)   & 0       & 1       & 1          \\
& Least FreqT(D)     & Freq Walks(Ob)   & 0       & 1       & 1          \\
& FreqT(D)           & Freq Walks(Ob)   & -0.5    & 1       & 1.5        \\
& FreqTypeT(D)       & Freq Walks(Ob)   & 0       & 1       & 1          \\ \cline{1-6}
Employer   & FreqT-counties(Ob) & Freq Walks(Ob)   & -1      & 0       & 1          \\ \cline{1-6}
Fed        & FreqT-counties(Ob) & International(D) & -0.5    & 1       & 1.5        \\ \cline{1-6}
Local gov  & Freq Walks(D)      & Home(D)          & 0       & 1       & 1          \\
& Visits(D)          & Home(D)          & 0       & 1       & 1         \\  \cline{1-6}
\end{tabularx}
\label{tab:kw_actor_feature_BIG}
\end{table}

\begin{table}[]
\centering
\footnotesize
\caption{KW with Dunn test to identify significantly different medians when obfuscated features are used for different actors and purposes. Median differences are statistically different (pval <0.01). Kruskal wallis chi-sq statistic for interaction(actor,privacy) = 192.84 pval<0.01; interaction(purpose,privacy) = 120.06 pval<0.01; interaction(feature,privacy) = 37.26 pval<0.01}
\begin{tabularx}{\linewidth}{|p{0.21\linewidth}|p{0.18\linewidth}|p{0.18\linewidth}|p{0.03\linewidth}|p{0.03\linewidth}|p{0.09\linewidth}|}
\cline{1-6}
\textbf{Component} & \textbf{Feature}    & \textbf{Feature}  & M1 & M2 & \text{\footnotesize{M1-M2}} \\ 
\cline{1-6}
Family                           & Obfuscated & Detailed & 1       & 0       & 1          \\ \cline{1-6}
Control diseases                 & Obfuscated & Detailed & 1       & 0       & 1         \\
\cline{1-6}
\end{tabularx}
\label{tab:kw_ap_obfuscate_BIG}
\end{table}

\begin{table}[ht]
\centering
\footnotesize
\caption{
KW with Dunn test to identify significantly different medians for Actors when different ethnicity/race participants are involved. Median differences are statistically different (pvalue <0.01). Kruskal wallis chi-sq statistic for interaction(actor,ethnicity) = 241.47 pval<0.01}
\begin{tabular}{|p{0.15\linewidth}|p{0.19\linewidth}|p{0.22\linewidth}|p{0.03\linewidth}|p{0.03\linewidth}|p{0.08\linewidth}|}
\hline
Ethnicity & Actor1            & Actor2               & M1 & M2 & \text{\footnotesize{M2-M1}} \\ \hline
Asian     & Employer          & Family               & -1 & 1  & 2     \\
          & Fed               & Family               & 0  & 1  & 1     \\ \hline
White     & Commercial        & Researchers          & 0  & 1  & 1     \\
          &                   & 911                  & 0  & 1  & 1     \\ \cline{3-6}
          
          & Employer          & Researchers          & -1 & 1  & 2     \\
          &                   & Commercial           & -1 & 0  & 1     \\
          &                   & Doc                  & -1 & 0  & 1     \\
          &                   & 911                  & -1 & 1  & 2     \\
          &                   & Family               & -1 & 0  & 1     \\
          &                   & Fed                  & -1 & 0  & 1     \\
          &                   & Law EA               & -1 & 0  & 1     \\
          &                   & Local gov            & -1 & 0  & 1     \\ \cline{3-6}
          & Fed               & Researchers          & 0  & 1  & 1     \\
          &                   & 911                  & 0  & 1  & 1     \\ \cline{3-6}
          & Law EA            & Researchers          & 0  & 1  & 1     \\
          &                   & 911                  & 0  & 1  & 1     \\ \cline{3-6}
          & Local gov         & Researchers          & 0  & 1  & 1     \\
          &                   & 911                  & 0  & 1  & 1     \\ \hline

\end{tabular}
\label{tab:kw_ethnicity_act}
\end{table}

\begin{table}[ht]
\centering
\footnotesize
\caption{
KW with Dunn test to identify significantly different medians for ethnicity/racial group participants when different purpose, feature and privacy groups are involved. Median differences are statistically different (pvalue <0.01). Kruskal wallis chi-sq statistic for interaction(purpose,ethnicity) = 187.67 pval<0.01; interaction(feature,ethnicity) = 132.08 
pval<0.01; interaction(privacy,ethnicity) = 68.63 pval<0.01}
\begin{tabular}{|p{0.15\linewidth}|p{0.19\linewidth}|p{0.22\linewidth}|p{0.03\linewidth}|p{0.03\linewidth}|p{0.07\linewidth}|}
\hline
 Ethnicity         & Feature1          & Feature2             & M1 & M2 & \text{\footnotesize{M2-M1}} \\
 \hline
White     & FreqT(D)          & Freq Walks(Ob)       & 0  & 1  & 1     \\
          &                   &                      &    &    &       \\ \hline
          & Feature           & Feature              & M1 & M2 & \text{\footnotesize{M2-M1}} \\  \hline
Asian     & Detailed          & Obfuscated           & 0  & 1  & 1    \\ \hline
          & Purpose1          & Purpose2             & M1 & M2 & \text{\footnotesize{M2-M1}} \\ \hline
Asian     & Monitor mobility  & Infra (Buildings)    & 0  & 1  & 1     \\ \hline
Black     & Public transit    & Ads                  & 0  & 1  & 1     \\
          & Terrorist attacks & Ads                  & 0  & 1  & 1     \\
          & Monitor mobility  & Ads                  & 0  & 1  & 1     \\ \hline
Hispanic  & Public transit    & Infra (Walk/Cycling) & 0  & 1  & 1     \\
          &                   & Infra (Buildings)    & 0  & 1  & 1     \\
          &                   & Work                 & 0  & 1  & 1     \\ \cline{3-6}
          & Ads               & Criminal activity    & -1 & 1  & 2     \\
          &                   & Terrorist attacks    & -1 & 1  & 2     \\
          &                   & Control diseases     & -1 & 1  & 2     \\
          &                   & Infra (Walk/Cycling) & -1 & 1  & 2     \\
          &                   & Infra (Buildings)    & -1 & 1  & 2     \\
          &                   & Monitor mobility     & -1 & 1  & 2     \\
          &                   & Work                 & -1 & 1  & 2     \\
          &                   & Wellness             & -1 & 1  & 2     \\ \hline
White     & Criminal activity & Infra (Walk/Cycling) & 0  & 1  & 1     \\
          &                   & Infra (Buildings)    & 0  & 1  & 1     \\ \cline{3-6}
          & Terrorist attacks & Infra (Walk/Cycling) & 0  & 1  & 1     \\
          &                   & Infra (Buildings)    & 0  & 1  & 1     \\ \cline{3-6}
          & Control diseases  & Infra (Buildings)    & 0  & 1  & 1     \\  \cline{3-6}
          & Monitor mobility  & Infra (Walk/Cycling) & 0  & 1  & 1     \\ 
          &                   & Infra (Buildings)    & 0  & 1  & 1     \\ \cline{3-6}
          & Work              & Infra (Walk/Cycling) & 0  & 1  & 1     \\
          &                   & Infra (Buildings)    & 0  & 1  & 1     \\ \cline{3-6}
          & Wellness          & Infra (Walk/Cycling) & 0  & 1  & 1     \\
          &                   & Infra (Buildings)    & 0  & 1  & 1     \\ \cline{3-6}
          & Ads               & Infra (Buildings)    & 0  & 1  & 1     \\
          &                   &                      &    &    &       \\  \hline
\end{tabular}
\label{tab:kw_ethnicity_p_f_priv}
\end{table}

\begin{table}[]
\centering
\footnotesize
\caption{
Subsample of KW with Dunn test to identify significantly different medians for actors, purposes and features when different education groups are involved. Median differences are statistically different (pvalue <0.01). Kruskal wallis chi-sq statistic for interaction(actor,education) = 198.53 pval<0.01; interaction(purpose,education) = 121.38 pval<0.01; interaction(feature,education) = 86.08 
pval<0.01; interaction(privacy,education) = 31.44 pval<0.01}
\begin{tabularx}{\linewidth}{|p{0.15\linewidth}|p{0.19\linewidth}|p{0.22\linewidth}|p{0.03\linewidth}|p{0.03\linewidth}|p{0.07\linewidth}|}
\cline{1-6}
Education & Actor1               & Actor2                  & M1 & M2 & \text{\footnotesize{M2-M1}} \\ \cline{1-6}
Bachelors and above                    & Employer          & Researchers          & -1      & 1       & 2          \\
&                      & Commercial           & -1      & 0       & 1          \\
&                      & Doc                  & -1      & 1       & 2          \\
&                      & 911                  & -1      & 1       & 2          \\
&                      & Family               & -1      & 1       & 2          \\
&                      & Fed                  & -1      & 0       & 1          \\
&                      & Law EA               & -1      & 1       & 2          \\ 
&                      & Local gov            & -1      & 1       & 2          \\
 \cline{1-6}
Highschool to Bachelor                & Employer          & Researchers          & -1      & 1       & 2          \\
&                      & Commercial           & -1      & 0       & 1          \\
&                      & Doc                  & -1      & 0       & 1          \\
&                      & 911                  & -1      & 1       & 2          \\
&                      & Family               & -1      & 0       & 1          \\
&                      & Fed                  & -1      & 0       & 1          \\
&                      & Law EA               & -1      & 0       & 1          \\
&                      & Local gov            & -1      & 0       & 1          \\  \cline{3-6}
& Fed               & Researchers          & 0       & 1       & 1          \\
&                      & 911                  & 0       & 1       & 1          \\  \cline{3-6}
& Law EA            & Researchers          & 0       & 1       & 1          \\
&                      & 911                  & 0       & 1       & 1          \\  \cline{3-6}
& Local gov         & Researchers          & 0       & 1       & 1          \\
&                      & 911                  & 0       & 1       & 1          \\
&                      &                         &         &         &            \\  \cline{1-6}
& Feature1             & Feature2                & M1 & M2 & \text{\footnotesize{M2-M1}} \\ \cline{1-6}
Bachelors and above                    & Least FreqT(D)    & Freq Walks(Ob)       & 0       & 1       & 1          \\
& FreqT(D)          & Freq Walks(Ob)       & 0       & 1       & 1          \\  \cline{1-6}
Highschool to Bachelor                 & FreqT(D)          & Freq Walks(Ob)       & 0       & 1       & 1          \\
&                      &                         &         &         &            \\  \cline{1-6}
& Purpose1             & Purpose2                & M1 & M2 & \text{\footnotesize{M2-M1}} \\ \cline{1-6}
Bachelors and above                    & Monitor mobility  & Public transit       & 0       & 1       & 1          \\
&                      & Criminal activity    & 0       & 1       & 1          \\
&                      & Control diseases     & 0       & 1       & 1          \\
&                      & Infra (Walk/Cycling) & 0       & 1       & 1          \\
&                      & Infra (Buildings)    & 0       & 1       & 1          \\ \cline{3-6}
& Work              & Infra (Walk/Cycling) & 0       & 1       & 1          \\
&                      & Infra (Buildings)    & 0       & 1       & 1          \\
 \cline{1-6}
& Feature              & Feature                 & M1 & M2 & \text{\footnotesize{M2-M1}} \\ \cline{1-6}
Bachelors and above                    & Detailed             & Obfuscated              & 0       & 1       & \\ \cline{1-6}
\end{tabularx}
\label{tab:kw_bachelors_apf}
\end{table}

\section{Survey and User Interface}
\label{section:surveyUI}
In this section, we discuss the survey in detail and the different sections in it. We built a custom survey webpage for better customization and functionality. Participants were shown ads for our survey on the Cint portal and they were directed to our survey page upon clicking on it.

The survey is divided into the following sections:
\begin{itemize}
    \item Consent and Briefing
    \item Demographic Information and General Technical Knowledge
    \item Understanding Level of comfort with location data sharing: Vignettes
    \item Privacy Attitudes
\end{itemize}


\subsection{Consent and Briefing}
\label{subsection:consentbrief}
\begin{figure*}
\centering
\includegraphics[width=0.7\linewidth]{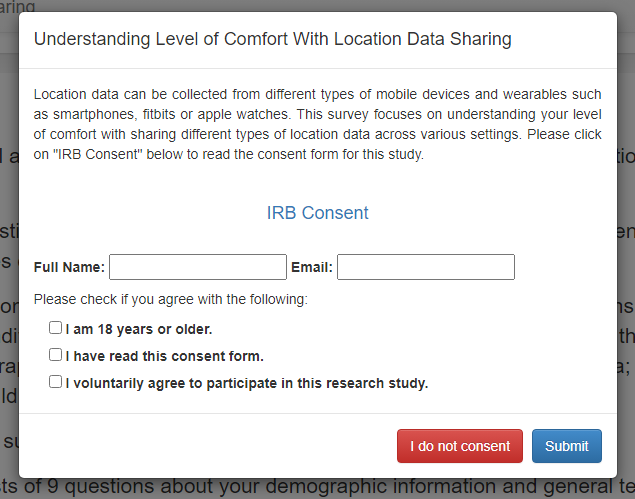}
\caption{Popup showing the consent page and Internal Review Board (IRB) document link}
\label{fig:irbpopup}
\end{figure*}

\begin{figure*}
\centering
\includegraphics[width=0.7\textwidth]{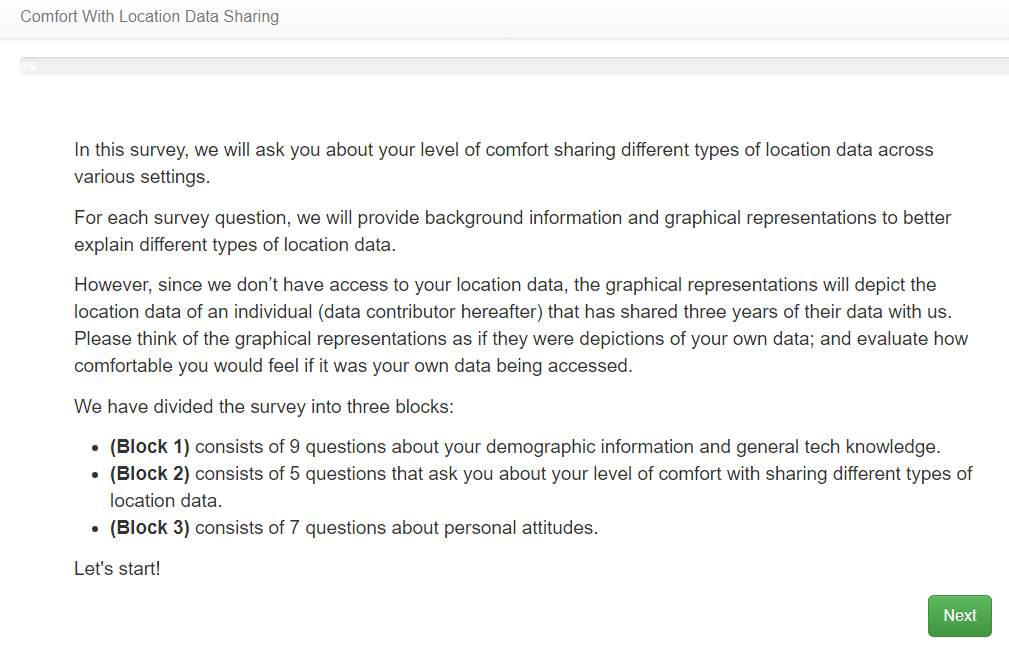}
\caption{The Introduction page}
\label{fig:consentbrief}
\end{figure*}

 Figure \ref{fig:consentbrief} is the landing page of our survey. The landing page opens with a pop-up window (see Figure \ref{fig:irbpopup}) briefly describing the objective of the study. The participants are also provided with a link to the Internal Review Board's (IRB) document detailing the project, ethical considerations, data handling and contact information. All participants are asked for their name, email, whether they are 18 years or older, consent to participate and whether they are voluntarily agreeing to participate in the survey. If they press the "I do not consent" button, they are directed away from the website. When the participant presses on the "submit" button, the pop-up window disappears and they can read the introduction page (Figure \ref{fig:consentbrief}) which details our study, the different sections and the number of questions, asking participants to think about the data and visualizations they see as their own. The participant moves to the next section when they click on the "Next" button. 


\subsection{Demographic Information and General Technical Knowledge}
\begin{figure*}
\centering
\includegraphics[width=0.7\textwidth]{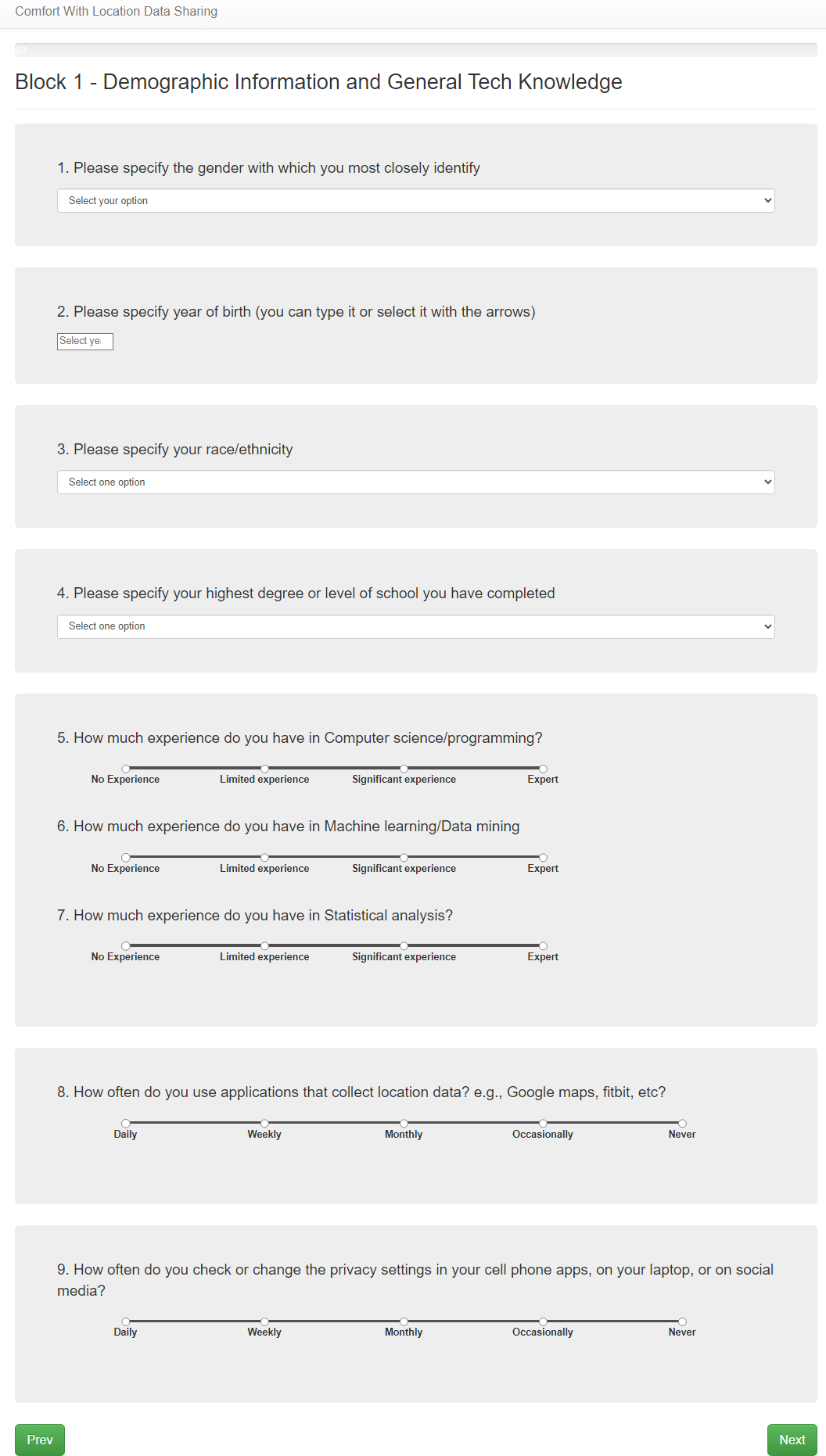}
\caption{Demographic and technical knowledge questions as seen by the survey participants}
\label{fig:demographic_page}
\end{figure*}

After reading the brief and moving to the next page, participants are asked about their demographic details and general awareness with different technologies (see Figure \ref{fig:demographic_page}).

\subsection{Understanding Level of comfort with location data sharing: Vignettes }

\begin{figure*}[ht]
\centering
\includegraphics[width=0.7\textwidth]{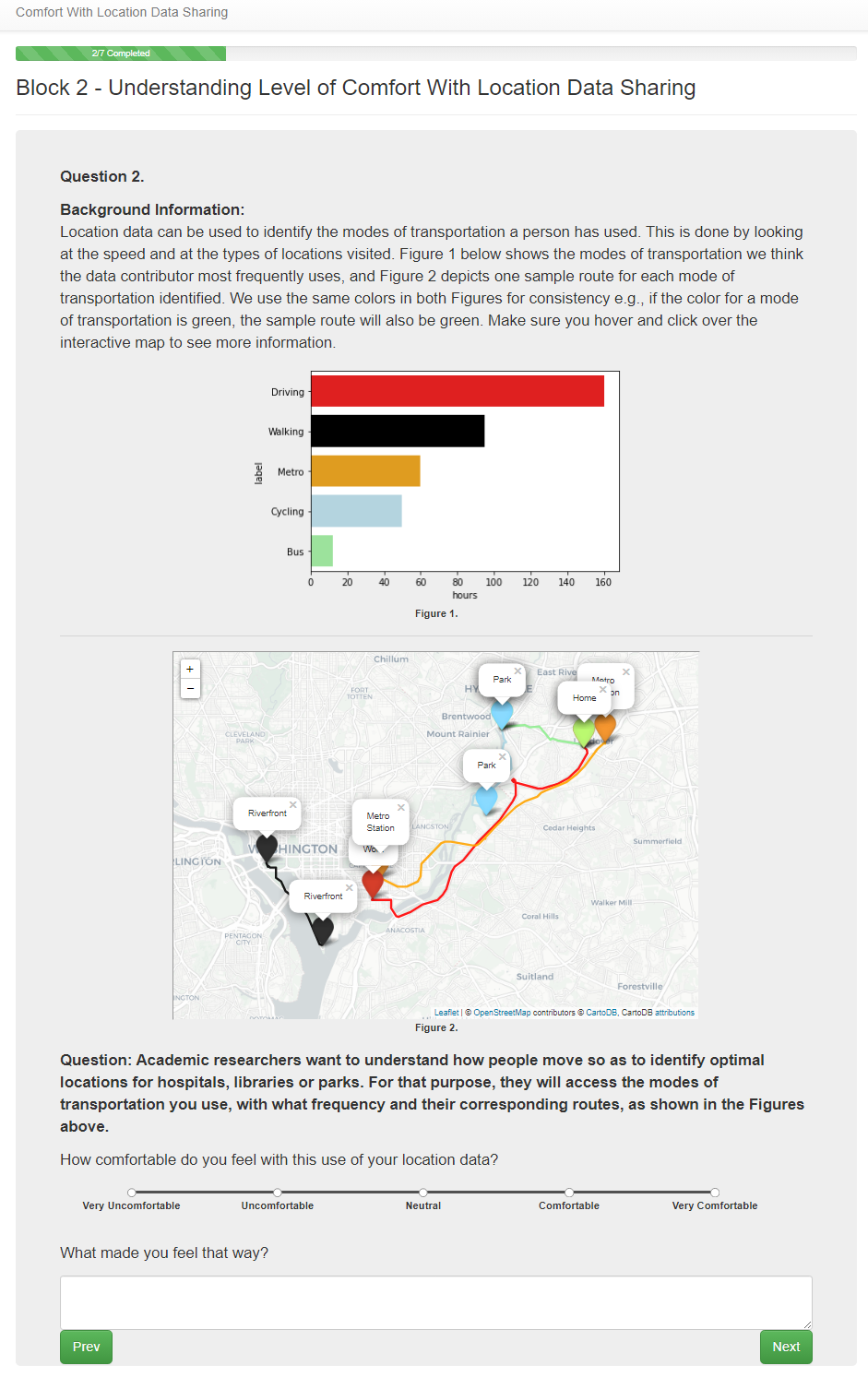}
\caption{Example of Vignette Question (Actor: Law enforcement agency- like city police department or a county sheriff's office, Purpose: Understand criminal activity, Feature: Places you visit). The feature is detailed and shows an interactive map with information}
\label{fig:locationdataqsdetailed}
\end{figure*}

\begin{figure*}[ht]
\centering
\includegraphics[width=0.7\textwidth]{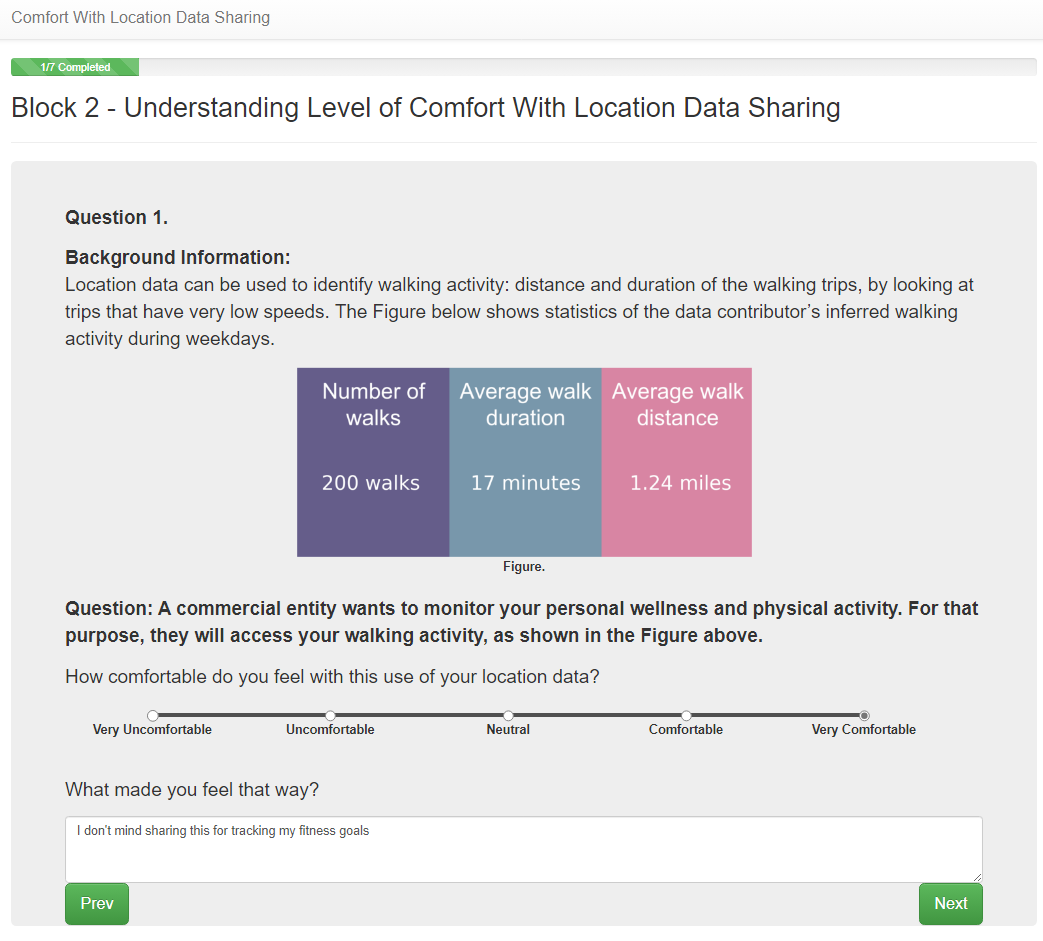}
\caption{Example of Vignette Question (Actor:Commercial entity, Purpose: Personal wellness and physical activity, Feature: Walking activity). The feature is obfuscating and only shows general statistics}
\label{fig:locationdataqspp}
\end{figure*}

There are 5 vignette questions asked to each participant. The questions are generated randomly from a set of questions and displayed on the participant's browser. There are two types of location features (Detailed vs Obfuscated). Fig. \ref{fig:locationdataqsdetailed} shows an example of a vignette focused on a detailed feature. In this example, participants see a chart showing statistics about the types of transportation inferred from location data as well as an interactive map with the detailed trajectories for each mode of transportation together with labels characterizing origin and destination place.
Participants can interact with the map to understand better all the information being shared. 
On the other hand, Figure \ref{fig:locationdataqspp} shows an example of a location feature computed in a privacy-preserving way. The participant can see general statistics about their walking activity and assess their level of comfort by selecting one of the five options. As can be observed, all vignettes also have a text box for participants to explain the rationale behind their choice. 

\subsection{Privacy Attitudes}
\label{martinAppendix}

\begin{figure*}
\centering
\includegraphics[width=0.5\textwidth]{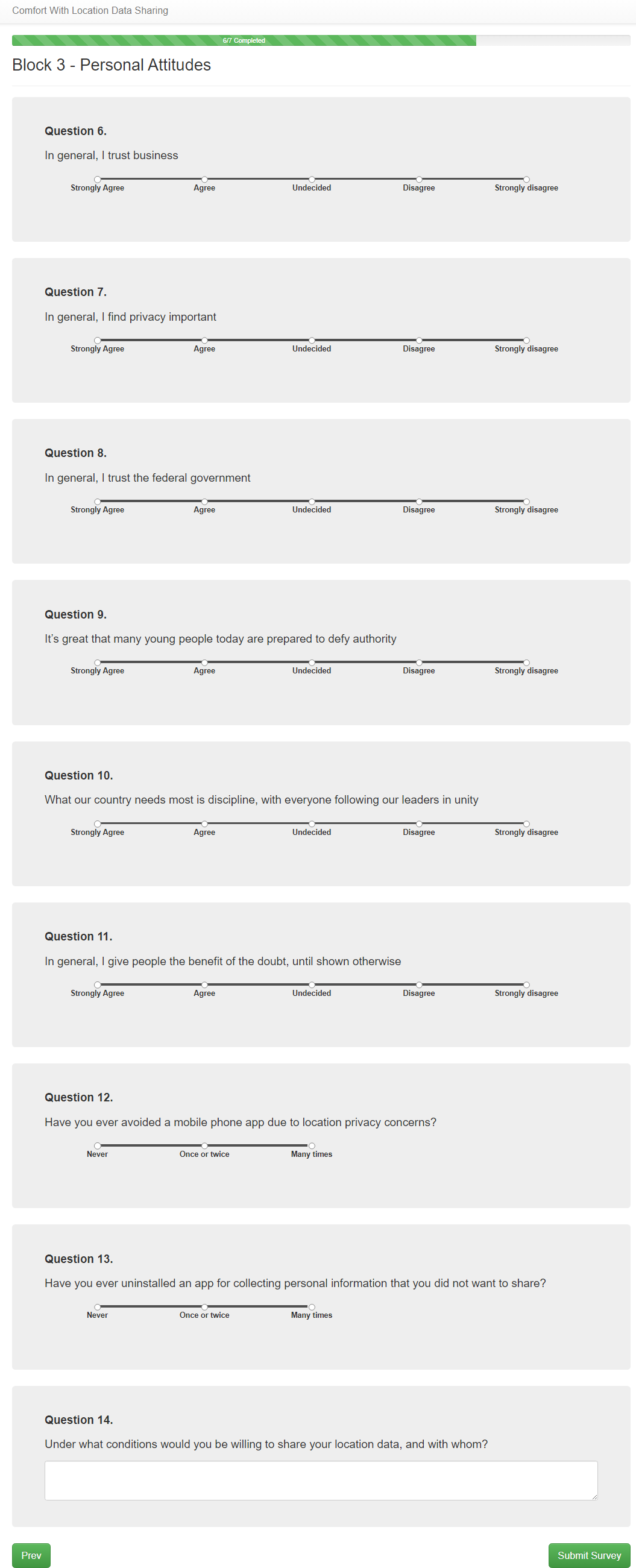}
\caption{Privacy attitude questions as seen by the survey participants}
\label{fig:attitude_webpage}
\end{figure*}

After completing the 5 vignette questions, participants are required to fill out the privacy attitudes survey, shown in Figure \ref{fig:attitude_webpage}. Studies like \cite{ICADataCollection,naeini,martin,colombia_privacy_study} have asked privacy and attitude questions which try to understand the person's general attitudes towards data sharing and privacy. We include the questions in Martin et. al. \cite{martin} as additional control variables in our regression to understand how different attitudes relate to comfort in sharing specific location features. These questions indirectly measure control, awareness and collection aspects as studied in Internet Users Information Privacy Concerns (IUIPC) \cite{IUIPC} based studies. As our goals are strictly understanding the broader horizon of location features and comfort, we focus our attention to these questions rather than forming the hierarchical IUIPC scores. We incorporate responses to the questions as ordinal variables (see response categories in \ref{tab:attitudes_responses} ). Table \ref{tab:attitude_privacy_coeffs} highlights questions which were significantly related to comfort levels. We observe similar trends to \cite{martin} in these questions. People with positive attitudes towards government and businesses show higher comfort levels. While people with higher defiance to authority had inverse relation with sharing location data. Two interesting findings: (1) Positive relation between computer science knowledge and comfort in sharing location data. (2) People who are more aware of their privacy settings were less comfortable in sharing location data.


\begin{figure}
\includegraphics[width=0.5\textwidth]{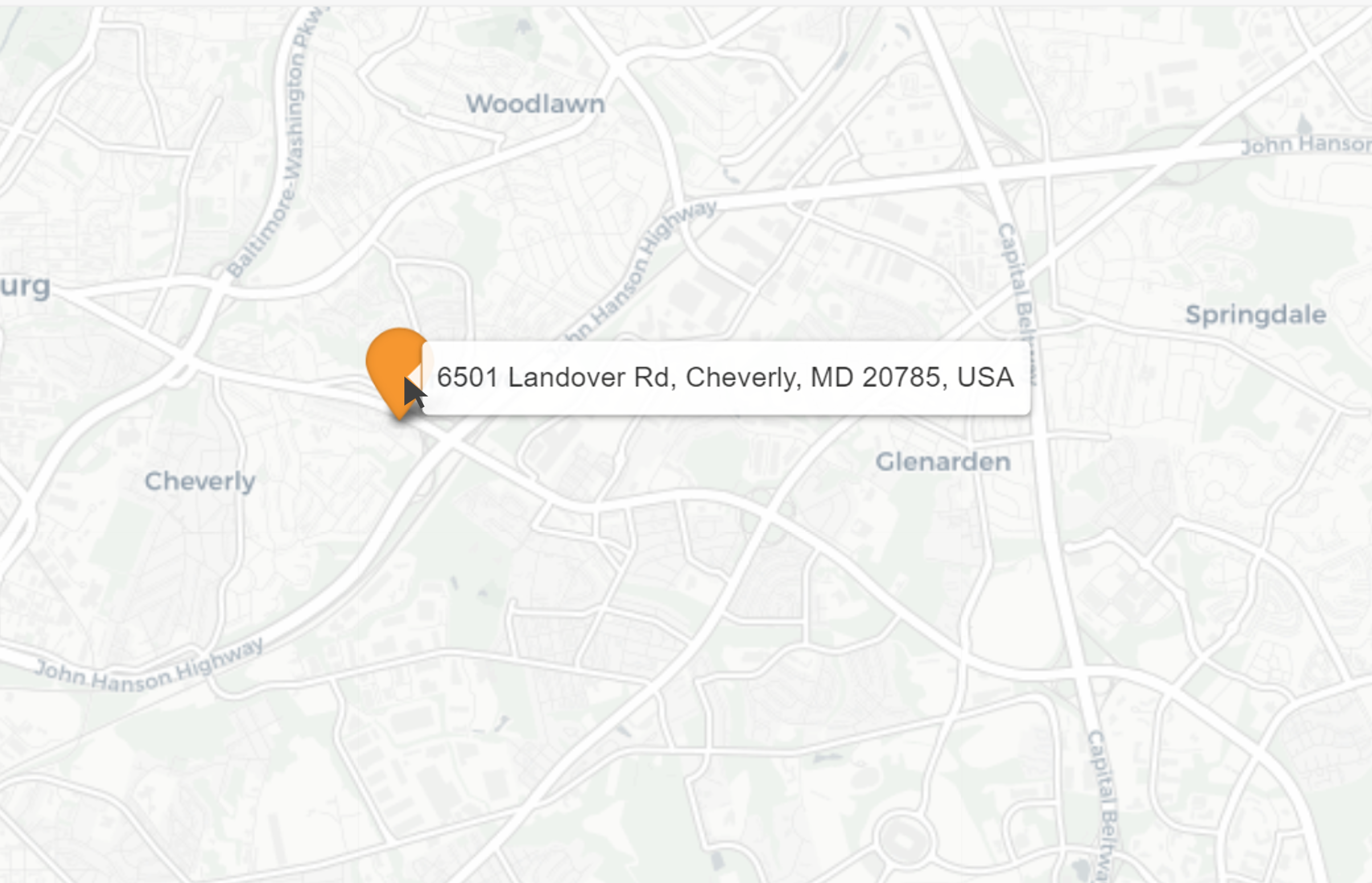}
\caption{Snapshot of interactive map with home Feature (Home(D)) as visible to participants. It shows the exact home location of the hypothetical data contributor using a pin marker.}
\label{fig:homeD}
\end{figure}
\begin{figure}
\includegraphics[width=0.5\textwidth]{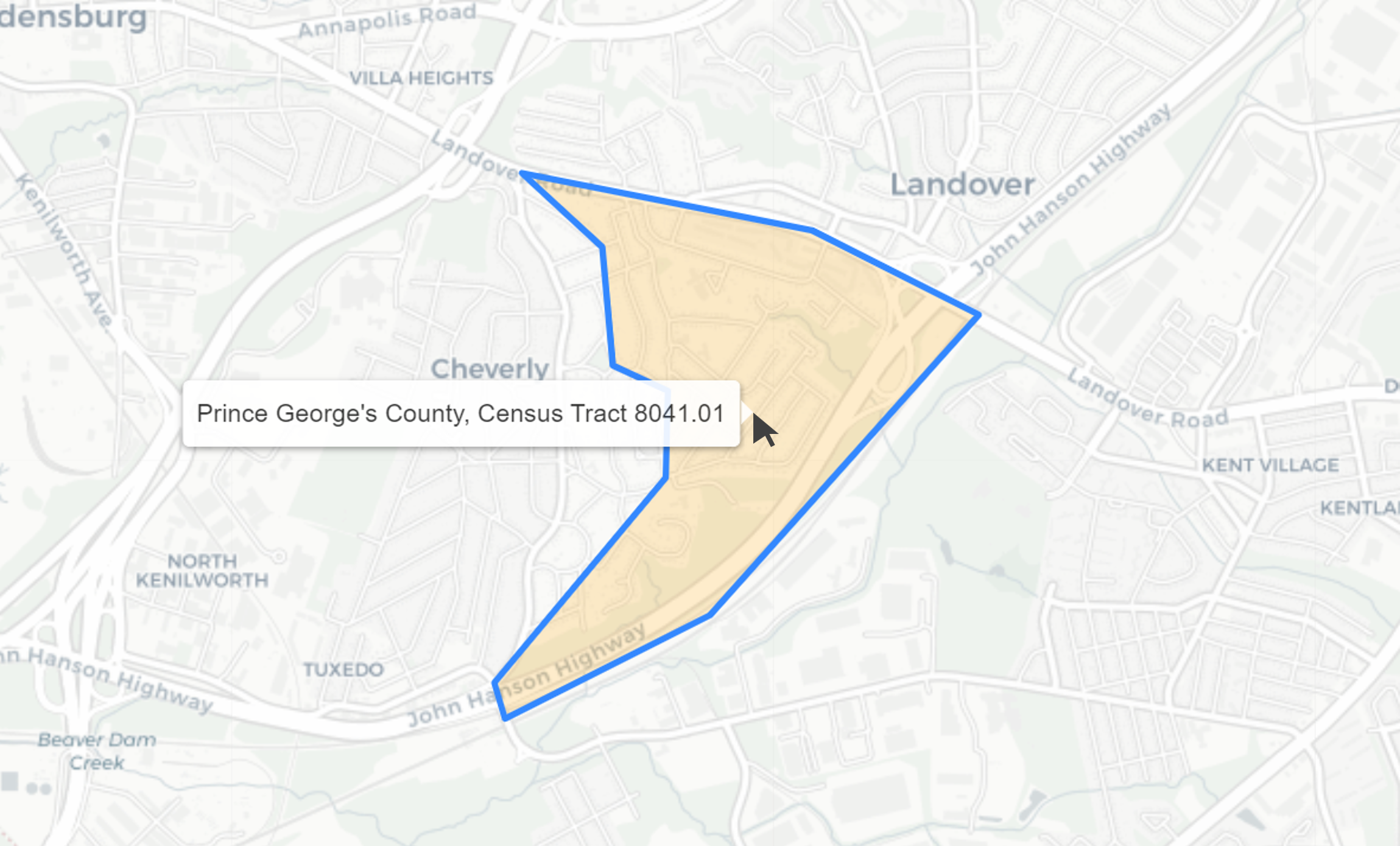}
\caption{Snapshot of interactive map with home Feature (Home(Ob)) as visible to participants. It shows the census tract region the hypothetical data contributor lives in.}
\label{fig:homePP}
\end{figure}
\begin{figure}
\includegraphics[width=0.5\textwidth]{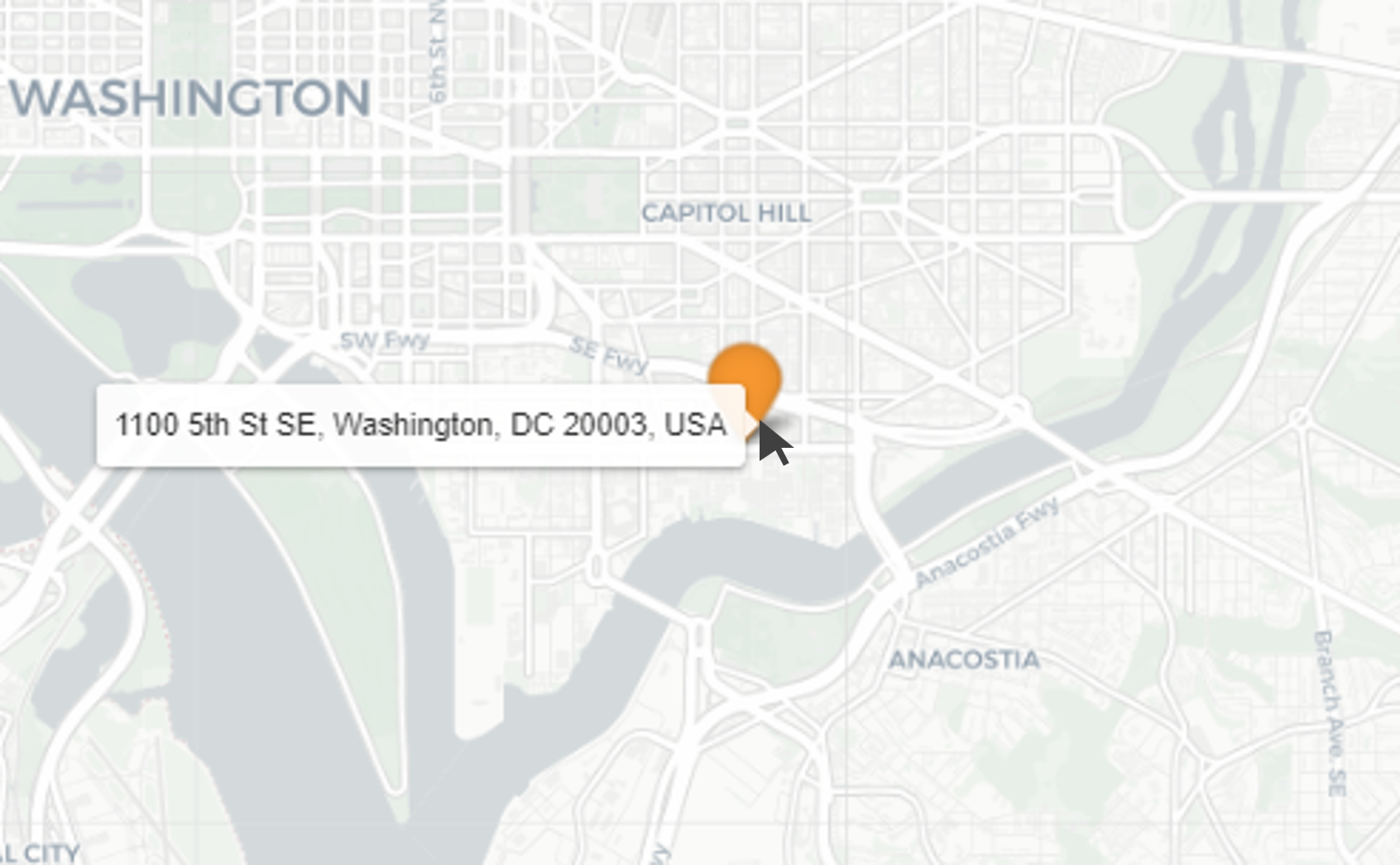}
\caption{Snapshot of interactive map with Work location feature (Work(D)) as visible to participants. It shows the exact work location of the hypothetical data contributor using a pin marker.}
\label{fig:workD}
\end{figure}
\begin{figure}
\includegraphics[width=0.5\textwidth]{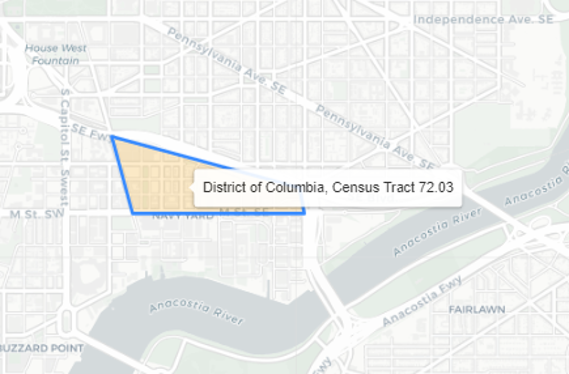}
\caption{Snapshot of interactive map with Work  feature (Work(Ob)) as visible to participants. It shows the census tract region where the hypothetical data contributor works at.}
\label{fig:workPP}
\end{figure}
\begin{figure}
\includegraphics[width=0.5\textwidth]{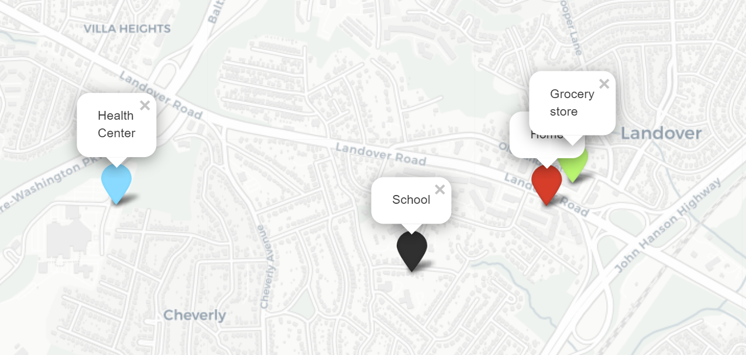}
\caption{Snapshot of interactive map with most Places of Visit (D) feature (Visits(D) as visible to participants. It shows the specific places the hypothetical data contributor frequently visits. The types of places are color coded (and consistent between the map and chart in Fig. \ref{fig:POVPP}) and labelled appropriately }
\label{fig:POVD}
\end{figure}
\begin{figure}
\includegraphics[width=0.5\textwidth]{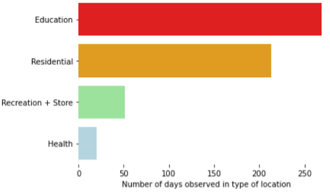}
\caption{Chart with Places of Visit (Ob) feature (Visits(Ob)) as visible to participants. The bar chart shows the different places of interest our hypothetical data contributor visits and the number of days such visits happen.}
\label{fig:POVPP}
\end{figure}
\begin{figure}
\includegraphics[width=0.5\textwidth]{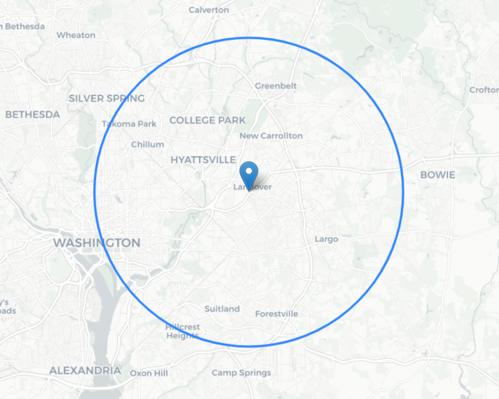}
\caption{Snapshot of interactive map with most Geographical area you spend most time feature (Radius(Ob) as visible to participants. The circle describes the region our hypothetical data contributor stays in. The pin denotes the center of this region}
\label{fig:ROG}
\end{figure}
\begin{figure}
\includegraphics[width=0.5\textwidth]{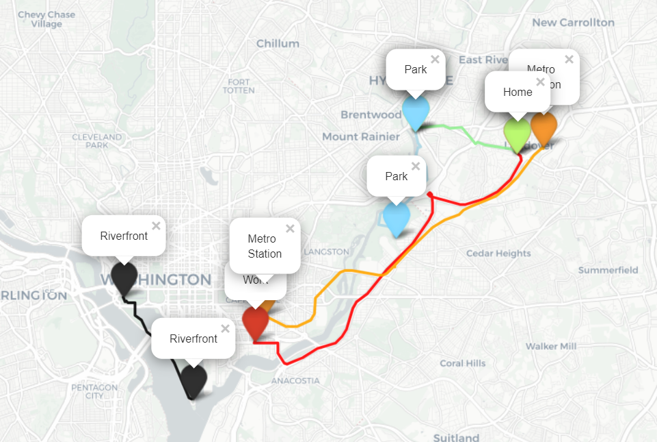}
\caption{Snapshot of interactive map with transportation Feature (Transportation(D)) as visible to participants. The map displays different modes of transport (driving, walking, etc) with different colors consistently between the map and the chart in fig. \ref{fig:modesPP}. For our hypothetical data contributor, the pins denote the start and end points and the line joining the pins denote the most frequent route taken by that mode of transport. E.g. Black pins and line segment denote the most frequent walking trip. The label "Riverfront" indicates the inferred type of trip}
\label{fig:modesD}
\end{figure}
\begin{figure}
\includegraphics[width=0.5\textwidth]{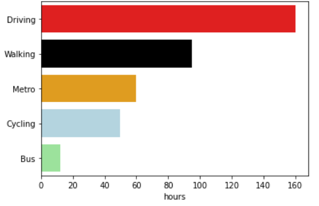}
\caption{Chart with modes of transportation Feature (Transportation(Ob)) as visible to participants. The bar chart shows the different modes of transport our hypothetical data contributor takes and the number of hours spent in each mode.}
\label{fig:modesPP}
\end{figure}

\begin{figure}
\includegraphics[width=0.5\textwidth]{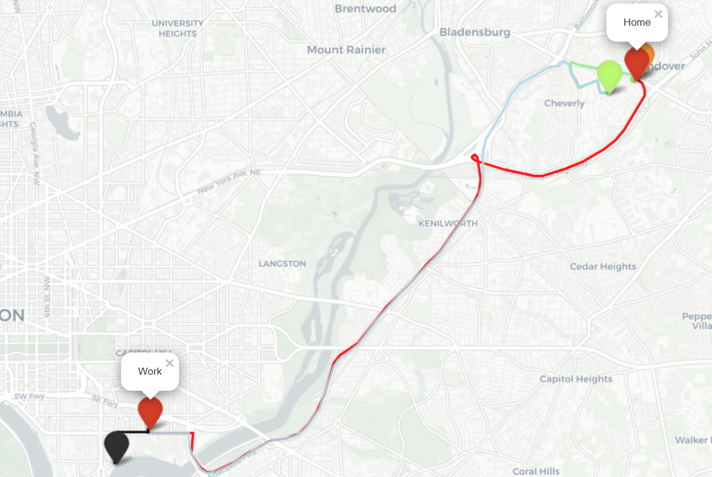}
\caption{Snapshot of interactive map with most frequent trips feature (FreqT (D) as visible to participants. For our hypothetical data contributor, the pins denote the start and end points and the line joining the pins denote one of the most frequent routes. E.g. red pins and line segment denote the most frequent trip between home and work.}
\label{fig:freqTD}
\end{figure}
\begin{figure}
\includegraphics[width=0.5\textwidth]{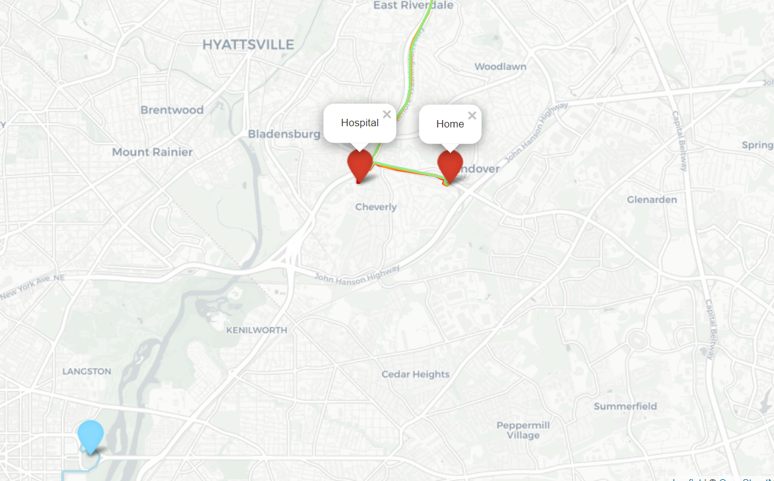}
\caption{Snapshot of interactive map with Least Frequent trips feature (LeastFreqT(D)) as visible to participants. For our hypothetical data contributor, the pins denote the start and end points and the line joining the pins denote one of the least frequent routes. E.g. red pins and line segment denote the least frequent trip between home and hospital.}
\label{fig:leastFreqT}
\end{figure}
\begin{figure}
\includegraphics[width=0.5\textwidth]{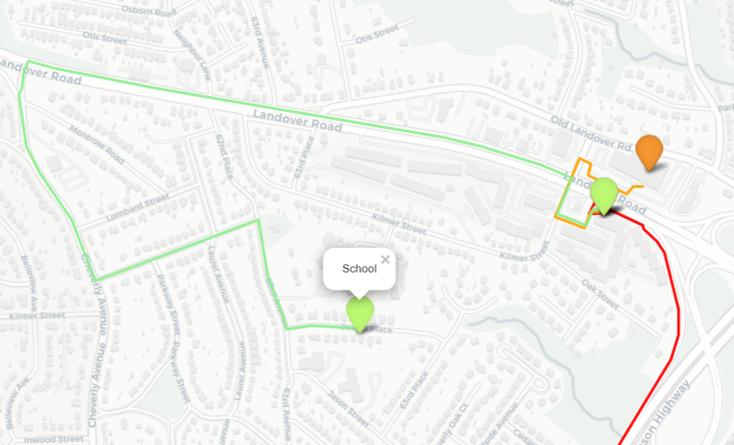}
\caption{Snapshot of interactive map with Most frequent type of trips Feature (Freq-TypeT(Ob)) as visible to participants. For our hypothetical data contributor, the pins denote the start and end points and the line joining the pins denote one of types of frequent trips. E.g. First, the most frequent trips are identified and labelled according to end points. Green pins and line segment denote a frequent trip to school which is between "home" and "school" and thus is a type of school trip.}
\label{fig:freqtypetrip}
\end{figure}
\begin{figure}
\includegraphics[width=0.5\textwidth]{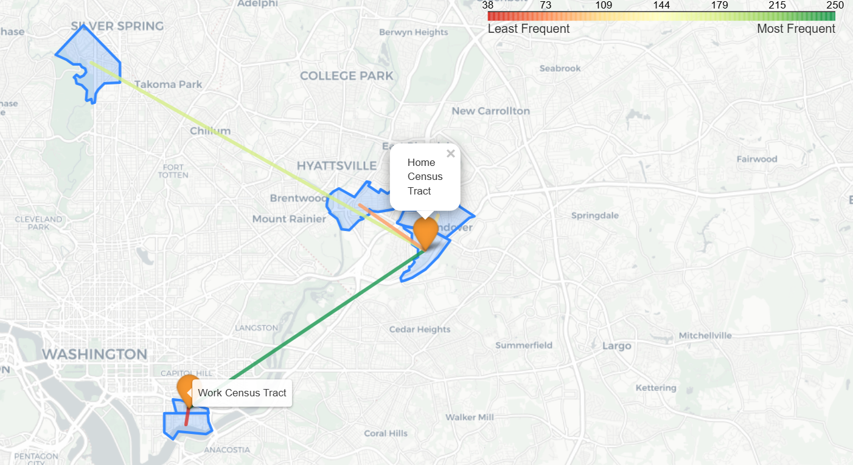}
\caption{Snapshot of interactive map with Most frequent trips represented by their origin and destination census tracts
and connected by a line Feature (FreqT-counties(Ob)) as visible to participants. We see that the home census tract and work census tract are joined by a straight green line (most frequent movement between these census tracts). }
\label{fig:FreqOD}
\end{figure}
\begin{figure}
\includegraphics[width=0.5\textwidth]{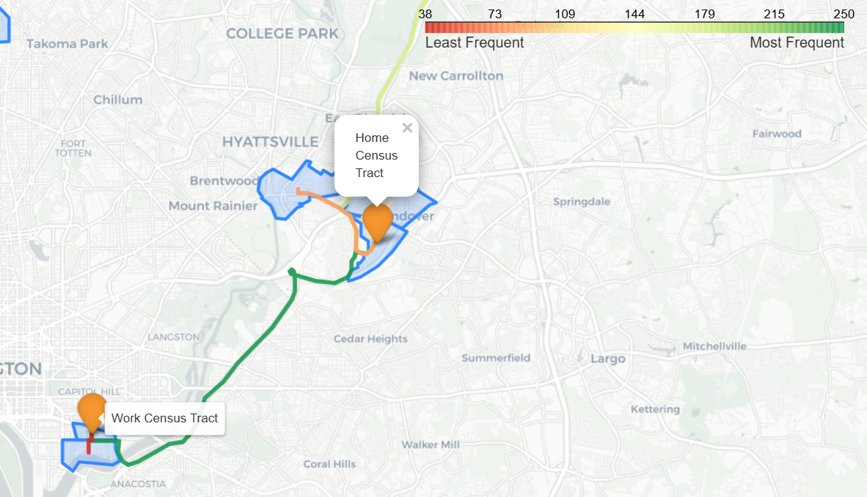}
\caption{Snapshot of interactive map with Most frequent trips represented by their origin and destination census tract and connected by a synthetic route extracted from Google Maps (FreqT-Google(Ob)) as visible to participants. The home census tract centroid and work census tract centroid are joined by a segment suggested by Google maps routes API .}
\label{fig:gmaps}
\end{figure}
\begin{figure}
\includegraphics[width=0.5\textwidth]{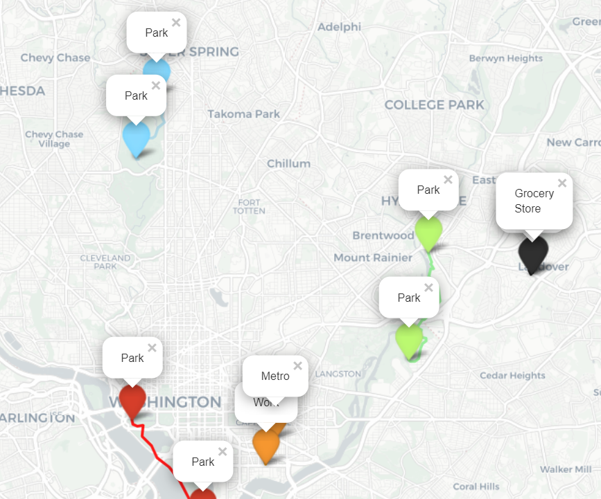}
\caption{Snapshot of interactive map with Your walking activity and corresponding routes feature (Freq Walks(D)) as visible to participants. For our hypothetical data contributor, the pins denote the start and end points and the line joining the pins denote the most frequent walking path taken. E.g. Red pins and line segment denote the most frequent walking trip. The label "park" on both pins indicates the start and end of the walk was in a park}
\label{fig:walkD}
\end{figure}
\begin{figure}
\includegraphics[width=0.5\textwidth]{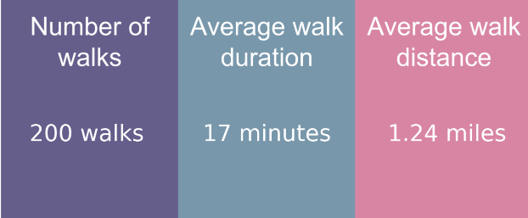}
\caption{Snapshot of the chart with Your walking activity (Freq Walks(Ob)) as visible to participants. Chart indicates frequency and duration of walks with no detailed trajectories}
\label{fig:walkPP}
\end{figure}
\begin{figure}
\includegraphics[width=0.5\textwidth]{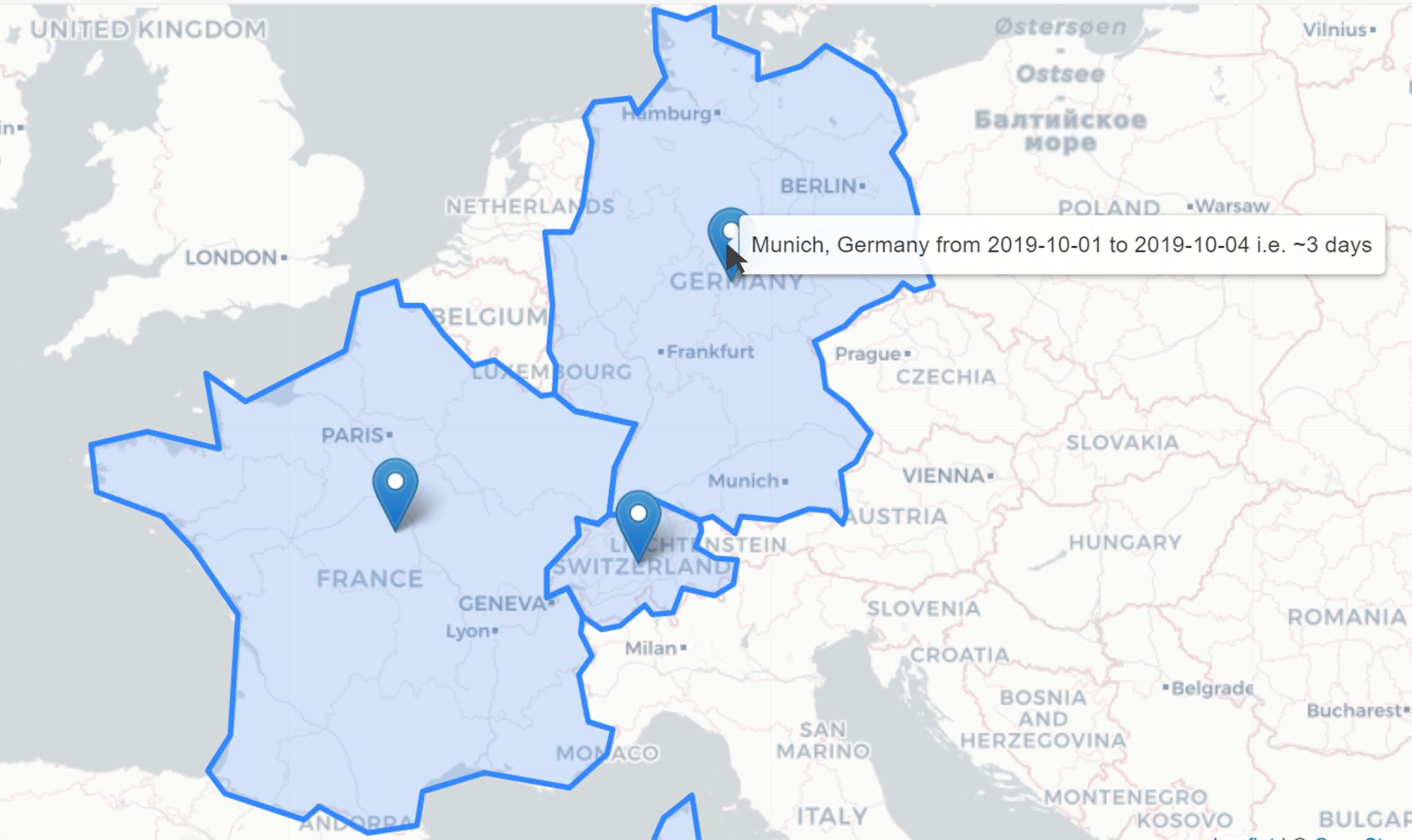}
\caption{Snapshot of interactive map with International visits feature (International(D)) as visible to participants. The map highlights the different countries visited and the pins indicate specific locations of visit with the duration of stay }
\label{fig:internationalD}
\end{figure}
\begin{figure}
\includegraphics[width=0.5\textwidth]{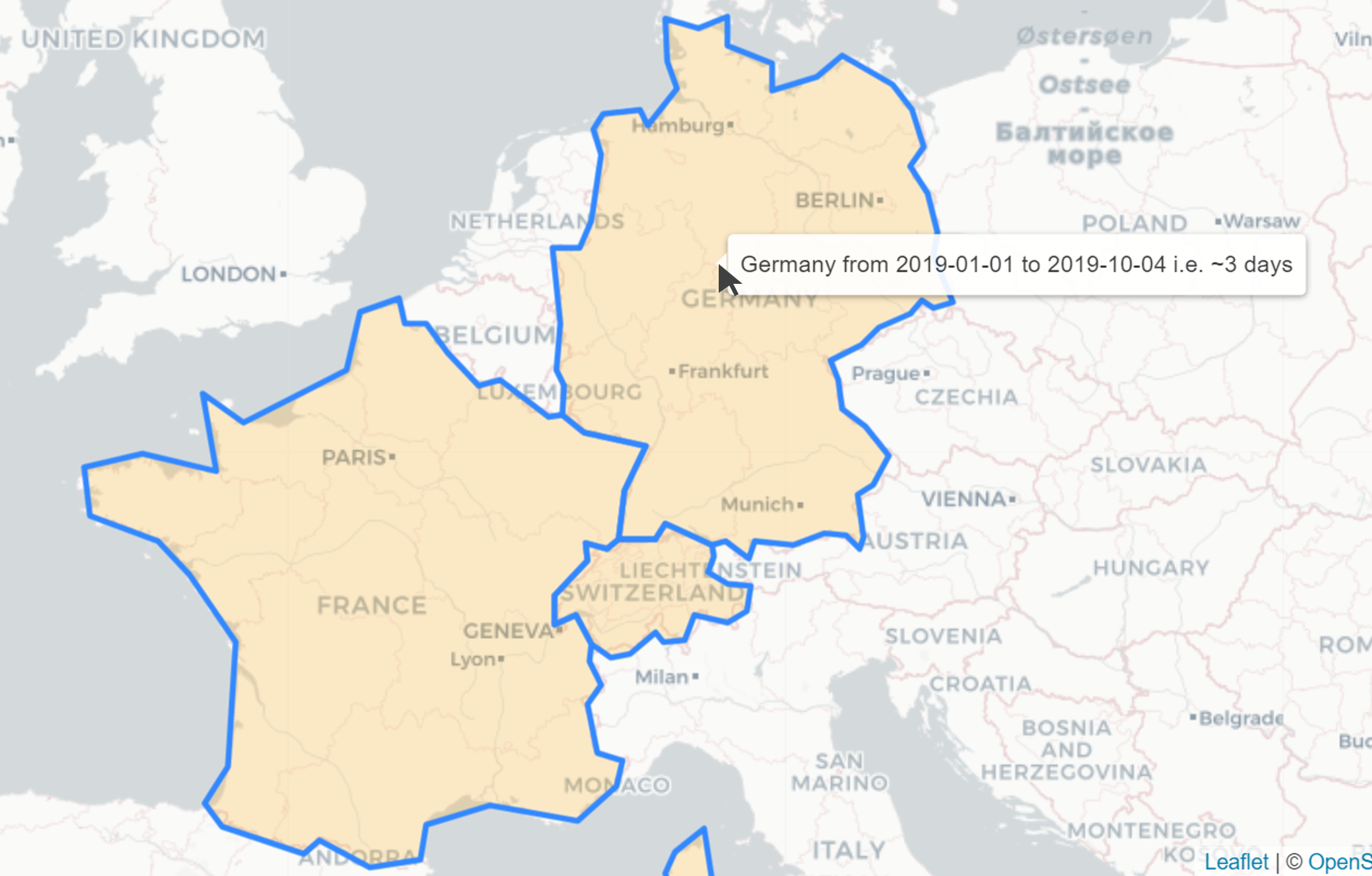}
\caption{Snapshot of interactive map with International visits feature (International(P)) as visible to participants. It specifies only the countries of visit and the duration of visit.}
\label{fig:internationalP}
\end{figure}

\end{document}